\documentclass{elsart}
\usepackage{amsmath}
\usepackage{wasysym}
\usepackage{fancyhdr}
\usepackage{fullpage}
\usepackage{amsbsy}
\usepackage{amssymb}
\usepackage{amscd}
\usepackage{amsfonts}
\usepackage{supertabular}
\usepackage{graphics}
\usepackage{verbatim}
\usepackage{subfigure}
\usepackage{epsfig}
\usepackage{xspace}
\usepackage{euscript}
\usepackage{alltt}
\usepackage{boxedminipage}
\usepackage{float}
\usepackage[colorlinks]{hyperref}
\usepackage{color}
\usepackage[all]{xy}
\usepackage[authoryear]{natbib}
\usepackage{t1enc}
\usepackage{times,exscale}
\usepackage{graphicx,calc}
\usepackage{pstricks,psfrag}
\usepackage{psfrag}
\usepackage{color}
\usepackage{subfigure}

\def\br{{\mathbf{r}}}

\def\bR{{\mathbf{R}}}

\begin{document}
\pagestyle{fancyplain}

    \lhead[\fancyplain{}{\sl Iyer, Radhakrishnan \& Gavini}]
          {\fancyplain{}{\sl Iyer, Radhakrishnan \& Gavini}}
    \rhead[\fancyplain{}
    {\sl \hfill }]
    {\fancyplain{}
    {\sl \hfill }}

\begin{frontmatter}
\title{Electronic-structure study of an edge dislocation in Aluminum and the role of macroscopic deformations on its energetics}
\author{Mrinal Iyer, Balachandran Radhakrishnan,}
\author{Vikram Gavini\corauthref{cor}}
\address{Department of Mechanical Engineering, University of Michigan, Ann Arbor, MI 48109, USA}
\corauth[cor]{Corresponding Author (\it vikramg@umich.edu)}
\begin{abstract}
We employed a real-space formulation of orbital-free density functional theory using finite-element basis to study the defect-core and energetics of an edge dislocation in Aluminum. Our study shows that the core-size of a perfect edge dislocation is around ten times the magnitude of the Burgers vector. This finding is contrary to the widely accepted notion that continuum descriptions of dislocation energetics are accurate beyond $\sim 1-3$ Burgers vector from the dislocation line. Consistent with prior electronic-structure studies, we find that the perfect edge dislocation dissociates into two Shockley partials with a partial separation distance of $12.8$~\AA. Interestingly, our study revealed a significant influence of macroscopic deformations on the core-energy of Shockley partials. We show that this dependence of the core-energy on macroscopic deformations results in an additional force on dislocations, beyond the Peach-Koehler force, that is proportional to strain gradients. Further, we demonstrate that this force from core-effects can be significant and can play an important role in governing the dislocation behavior in regions of inhomogeneous deformations.
\end{abstract}


\end{frontmatter}

\section{Introduction}
Dislocations are line defects in crystalline materials which play an important role in governing the deformation and failure mechanisms in solids. The energetics of dislocations and their interactions with other defects---solute atoms, precipitates, grain boundaries, surfaces and interfaces---significantly influence the mechanical properties of crystalline materials (cf.~e.g.~\citet{Hirth, Meyers_review,Uchic2004,Trinkle2005,Zhu2007,Gavini3,Curtin2010}). For instance, the kinetic barriers for dislocation motion---dislocation glide and climb---and their dependence on crystallographic planes and directions govern ductility and creep in metals~\citep{Bulatov1995,Duesbery1998,Lu2000,VanVliet2010}. Interaction of dislocations with vacancies, solute atoms and precipitates results in solid-solution strengthening/softening, precipitate hardening and aging in metals~\citep{Pollock1992,Lu2002,Trinkle2005,Yasi2010,Curtin2010}. Further, dislocation interactions with grain boundaries and surfaces are responsible for the observed strengthening mechanisms like the Hall-Petch effect~\citep{Hall-Petch2004}, and enhanced yield strength in surface dominated nanostructures~\citep{Uchic2004,Greer2006}.

The behavior of dislocations (nucleation, kinetics, evolution) in crystalline materials is governed by physics on multiple length-scales. In particular, a dislocation produces elastic fields that are long-ranged, and through these elastic fields interacts with other defects and external loads at macroscopic scales. On the other hand, the quantum-mechanical and atomistic scale interactions play an important role in governing the nucleation and kinetics of these defects. While atomistic scale interactions can significantly influence the behavior of dislocations, these are localized to a region around the dislocation line referred to as the dislocation-core. Thus, the energy of a dislocation comprises of the stored elastic energy ($E_{elastic}$), associated with the elastic fields outside the dislocation-core, and a core-energy ($E_{core}$) associated with quantum-mechanical and atomistic scale interactions inside the dislocation-core. Continuum theories based on elastic formulations have been widely used to study deformation and failure mechanisms mediated through dislocations (cf. e.~g.~\citet{Rice1992,Fleck1994,Nix1998,Ghoniem2000,Arsenlis2002}), where the energetics of dislocations are solely determined by the elastic energy and the core-energy is often assumed to be an inconsequential constant. In order to overcome the inability of continuum theories to describe the dislocation-core, explicit atomistic calculations based on empirical interatomic potentials have also been employed to study deformation mechanisms mediated by dislocations (cf. e.~g.~\citet{Tadmor1996,Hamilton1998,Gumbsch1999,Li2002,Cai2004,Marian2004}), and have provided many useful insights. However, interatomic potentials, whose parameters are often fit to bulk properties, may not accurately describe the defect-core which is governed by the electronic-structure (cf. e.~g.~\citet{Gumbsch1991,Arias2000,Trinkle2008}).

Electronic-structure calculations using plane-wave implementations of density functional theory (DFT) have been employed to study the dislocation core-structure in a wide range of crystalline materials (cf. e.~g.~\citet{Arias2000,Chrzan2000,Jacobsen2003,Trinkle2008,Willaime2009}) and the energetics of dislocation-solute interactions in metals with different crystallographic symmetries~\citep{Trinkle2005,Yasi2010}. As the displacement fields produced by isolated dislocations are not compatible with periodic boundary conditions, these calculations have either been restricted to artificial dipole and quadrapole configurations of dislocations or free-surfaces have been introduced to contain isolated dislocations. Recent efforts have also focused on the development of flexible boundary conditions by extending the lattice Green's function method to electronic-structure calculations~\citep{TrinklePRB2008}. Flexible boundary conditions accurately account for the long-ranged elastic fields of an isolated dislocation~\citep{Trinkle2008}, however, the electronic-structure in these studies is computed by introducing free surfaces to accommodate the restrictive periodic boundary conditions associated with plane-wave based DFT implementations. While these aforementioned studies have provided useful insights into the dislocation core-structure, a direct quantification of the dislocation core-energy solely from electronic-structure calculations and its role in governing dislocation behavior has remained elusive thus far. We note that some prior \emph{ab initio} studies using a dipole or quadrapole configuration of dislocations (cf. e.g.~\cite{Chrzan2000,Li2004,Willaime2009}), have attempted to indirectly compute the core-energy of an isolated dislocation by subtracting from the total energy the elastic interaction energy between dislocations in the simulation cell and their periodic images. This approach assumes the spacing between dislocations is large enough that the dislocation-cores do not overlap. However, these prior studies have been conducted on computational cells containing a few hundred atoms, which, as demonstrated in this work, are much smaller than the core-size of an isolated perfect edge dislocation in Aluminum. 

In this work, we conduct large-scale electronic-structure calculations using orbital-free density functional theory to study an edge dislocation in Aluminum. In our study, we use the WGC kinetic energy functional~\citep{Yan2} which has been shown to be in good agreement with Kohn-Sham DFT for a wide range of material properties in Aluminum~\citep{Yan2,Carter-Al-Mg,Ho,Carter2009}. We employ a local real-space formulation of orbital-free density functional theory~\citep{Gavini1,GaviniPRB2010}, where the extended interactions are reformulated as local variational problems. This real-space formulation of orbital-free density functional theory is used in conjunction with the finite-element basis that enables the consideration of complex geometries and general boundary conditions, which is crucial in resolving the aforementioned limitations of plane-wave basis in the study of energetics of isolated dislocations.

We begin our study by computing the size of the dislocation-core for a perfect edge dislocation in Aluminum. To this end, we consider a perfect edge dislocation with the atomic positions given by isotropic elasticity theory. For these fixed atomic positions, we identify the region where the perturbations in the electronic-structure arising from the defect-core are significant and have a non-trivial contribution to the dislocation energy. This allows us to unambiguously identify the dislocation-core from the viewpoint of energetics. Our study suggests that the dislocation core-size of a perfect edge dislocation is about $10|\mathbf{b}|$, where $\mathbf{b}$ denotes the Burgers vector. This estimate is much larger than conventional estimates based on displacement fields, which suggest a dislocation core-size of $1-3|\mathbf{b}|$~\citep{Hirth,Peierls1940,Kioussis2007,Weinberger2008}, and underscores the long-ranged nature of the perturbations in the electronic fields arising from defects. We note that a similar long-ranged nature of the electronic fields was observed in recent studies on point defects~\citep{Gavini2,GaviniPRB2010,GaviniJMPS2011}. As a next step in our study, we allow for atomic relaxations, and the perfect edge dislocation dissociates into Shockley partials with a partial separation distance of $12.8$~\AA. The dislocation energy per unit length of the relaxed Shockley partials in the simulation domain corresponding to the identified core-size, which denotes the dislocation core-energy, is computed to be $0.4$~$eV/\mbox{\AA}$.

We next study the role of macroscopic deformations on the dislocation core-energy and core-structure. In particular, we considered a wide range of macroscopic deformations including: (i) equi-triaxial strains representing volumetric deformations; (ii) uniaxial strains along the the Burgers vector, normal to the slip plane, and along the dislocation line; (ii) equi-biaxial strains along these crystallographic directions; (iii) shear strains along these crystallographic directions, excepting the one which results in dislocation glide. Interestingly, we find that the dislocation core-energy is significantly influenced by applied macroscopic deformations. In particular, we find that this core-energy dependence on macroscopic strains is non-linear and non-monotonic. Further, this core-energy dependence on macroscopic deformations suggests that the dislocation experiences an additional configurational force beyond the Peach-Koehler force. This additional force arising from core-effects, and referred to as the core-force, is proportional to the strain gradients and can be significant in regions of inhomogeneous deformations. In particular, we estimate that, in the case of a dislocation dipole with the two dislocations aligned at $45^{\circ}$ to the slip direction, the core-force has a glide component which is greater than the Peierls Nabarro force even at a distance of $25~nm$. While we observe a significant influence of macroscopic deformations on the core-energy, the core-structure, studied using differential displacement plots, did not change significantly in response to the applied macroscopic uniaxial, biaxial and triaxial strains. However, the shear strain which exerts a glide force on the screw components of the Shockley partials showed a significant influence, as expected, on the core-structure.

The remainder of this paper is organized as follows. Section~\ref{sec:OFDFT} presents an overview of the local real-space formulation of orbital-free DFT employed in this work. Section~\ref{sec:Results} presents our electronic-structure study of the edge dislocation in Aluminum, along with a discussion of the new findings and its implications. We finally conclude with an outlook in Section~\ref{sec:Conclusions}.

\section{Overview of orbital-free DFT}\label{sec:OFDFT}
In this section, for the sake of completeness and to keep the discussion self-contained, we provide an overview of the local real-space formulation of orbital-free DFT employed in the present work and the finite-element discretization of the formulation. We refer to~\citet{GaviniPRB2010,Motamarri-OFDFT2012} for a comprehensive discussion on these topics.

\subsection{Local real-space formulation of orbital-free DFT}\label{RS-OFDFT}
The ground-state energy in density functional theory~\citep{Martin} for a materials system containing $N$ nuclei and $N_e$ electrons is given by
\begin{equation}\label{eqn:Energy}
E(\rho,\bR) = T_{s}(\rho) + E_{xc}(\rho) + E_{H}(\rho) + E_{ext}(\rho,\bR) + E_{zz}(\bR)\,,
\end{equation}
where $\rho(\br)$ denotes the ground-state electron density and $\bR=\{\bR_1,\bR_2,\ldots,\bR_N\}$ denotes the vector collecting the positions of atoms. In the above, $T_s$ denotes the kinetic energy of non-interacting electrons; $E_{xc}$ denotes the exchange-correlation energy; $E_H$ denotes the Hartree energy or the electrostatic interaction energy between the electrons; $E_{ext}$ denotes the electrostatic interaction energy between the electrons and nuclei; and $E_{zz}$ denotes the nuclear-nuclear repulsion energy.

In the Kohn-Sham formalism, the kinetic energy of non-interacting electrons, $T_s$, is evaluated using the single-electron wavefunctions computed from the Kohn-Sham equations. In orbital-free DFT, in order to circumvent the cubic-scaling ($O(N_e^3)$) complexity in evaluating the single-electron wavefunctions or associated density matrix in the Kohn-Sham formalism, the kinetic energy of non-interacting electrons is modeled as an explicit functional of electron density~\citep{Parr}. This reduces the problem of evaluating the ground-state energy and electron density to a variational problem that scales linearly with system size. The widely used model for $T_s$, which has been shown to be a transferable model, in particular, for Al, Mg and Al-Mg materials systems~\citep{Carter-Al-Mg,Ho}, is the Wang-Govind-Carter (WGC) kinetic energy functional~\citep{Yan2} given by
\begin{equation}\label{eqn:KE}
T_s(\rho)=C_F\int \rho^{5/3}(\br)d\br + \frac{1}{2}\int |\nabla \sqrt{\rho(\br)}|^2 d\br + T^{\alpha,\beta}_{\rm K} (\rho) \,,
\end{equation}
where
\begin{equation}
T^{\alpha,\beta}_{\rm K}(\rho)=C_F\int\int \rho^{\alpha}(\br) K(|\br-\br'|; \rho(\br),\rho(\br')) \rho^{\beta}(\br') d\br d\br' \,\,.
\end{equation}
In equation~\eqref{eqn:KE}, the first term denotes the Thomas-Fermi energy with $C_F=\frac{3}{10}(3\pi^2)^{2/3}$, and the second term denotes the von-Weizs$\ddot{a}$cker correction~\citep{Parr}. The last term denotes the density dependent kernel energy, $T^{\alpha,\beta}_{\rm K}$, where the kernel $K$ and the parameters $\alpha$ and $\beta$ are chosen such that the linear response of a uniform electron gas is given by the Lindhard response. In the WGC functional, the parameters are chosen to be $\{\alpha,\beta\}=\{5/6+\sqrt{5}/6,5/6-\sqrt{5}/6\}$, and the density dependent kernel is expanded as a Taylor series about the bulk average electron density to second order, thus yielding a series of density independent kernels~\citep{Yan2}.

The second term in equation~\eqref{eqn:Energy}, $E_{xc}$, denoting the exchange-correlation energy, includes all the quantum-mechanical interactions and is modeled in density functional theory. Widely used models for the exchange-correlation energy, especially for solid-state calculations of ground-state properties, include the local density approximation (LDA) and the generalized gradient approximation (GGA)~\citep{Martin}, and have been demonstrated to be transferable models for a range of materials systems and properties. The last three terms in equation~\eqref{eqn:Energy} constitute the classical electrostatic energy between electrons and the nuclei. We note that all the terms in the energy functional~\eqref{eqn:Energy} are local, excepting the electrostatic energy and the kernel energy, which are extended in real-space. As a local real-space formulation is desirable for scalable numerical implementations as well as the development of coarse-graining techniques, a local reformulation of these extended interactions has been proposed and implemented recently~\citep{Gavini1,GaviniPRB2010}. By noting that the extended interactions in electrostatics are governed by the $\frac{1}{|\br-\br'|}$ kernel, which is the Green's function of the Laplace operator, the electrostatic interaction energy can be reformulated as the following local variational problem:
\begin{equation}\label{eqn:elReformulation}
E_H+E_{ext}+E_{zz}
= -\inf_{\phi \in \mathcal{Y}} \left\{\frac{1}{8\pi}\int |\nabla \phi(\br)|^2 d\br - \int \big(\rho(\br) + \sum_{I=1}^{N}b_{I}(\br,\bR_I)\big)\phi(\br)d\br\right\}-E_{self}\,,
\end{equation}
where $b_I(\br,\bR_I)$ denotes the charge distribution corresponding to the ionic pseudopotential of the $I^{th}$ nucleus, $\phi$ denotes the electrostatic potential corresponding to the total charge distribution comprising of the electrons and the nuclei, and $\mathcal{Y}$ denotes a suitable function space incorporating appropriate boundary conditions for the problem being solved. In the above, $E_{self}$ denotes the self energy of the nuclear charge distributions which can be evaluated, again, by taking recourse to the Poisson equation (cf.~\citet{Motamarri-OFDFT2012, Motamarri-KSDFT2013}).

A local reformulation of the extended interactions in the kernel kinetic energy functional has been proposed in~\citet{Choly2002}, and has been subsequently recast into a variational problem in~\citet{GaviniPRB2010}. It was demonstrated in~\citet{Choly2002} that the series of density independent kernels obtained from the Taylor series expansion of the WGC density dependent kernel can each be approximated in the Fourier-space using a sum of partial fractions. Using this approximation, a real-space reformulation for these extended interactions has been proposed by taking recourse to the solutions of a series of complex Helmholtz equations. In particular, if $\bar{K}(|\br-\br'|)$ denotes a density independent kernel, and is approximated in the Fourier-space using the approximation $\hat{\bar{K}}(\mathbf{q})\approx \sum_{j=1}^{m}\frac{A_j |\mathbf{q}|^2}{|\mathbf{q}|^2+B_j}$, the local reformulation of the kernel energy is given by the following saddle point problem~\citep{GaviniPRB2010,Motamarri-OFDFT2012}:
\begin{equation}\label{eqn:kerReformulation}
\int\int \rho^{\alpha}(\br) \bar{K}(|\br-\br'|) \rho^{\beta}(\br') d\br d\br'= \inf_{{\raisebox{-0.7ex}{$\scriptstyle\mathbf{\tilde{\omega}}_{\alpha}\in \mathcal{Z}$}}}\,\sup_{\mathrm{\tilde{\omega}}_{\beta} \in \mathcal{Z}}\bar{L}(\rho,\mathbf{\tilde{\omega}}_{\alpha},\mathbf{\tilde{\omega}}_{\beta})\,,
\end{equation}
where
\begin{equation}
\begin{split}
\bar{L}(\rho,\mathbf{\tilde{\omega}_{\alpha}},\mathbf{\tilde{\omega}_{\beta}}) = & \sum_{j=1}^{m}\Bigl\{\int \Bigl[\frac{1}{A_j\;B_j}\nabla\omega_{\alpha_j}(\br)\cdot\nabla\omega_{\beta_j}(\br) + \frac{1}{A_j}\omega_{\alpha_j}(\br)\omega_{\beta_j}(\br) \\
&+ \omega_{\beta_j}(\br)\rho^{\alpha}(\br) + \omega_{\alpha_j}(\br)\rho^{\beta}(\br) + A_J \rho^{(\alpha+\beta)}(\br)\Bigr]d\br\Bigr\}\,.
\end{split}
\end{equation}
In the above, $\tilde{\omega}_{\alpha}$ and $\tilde{\omega}_{\beta}$ denote the vector of potential fields given by $\mathbf{\tilde{\omega}}_{\alpha}=\{\omega_{\alpha_1},\omega_{\alpha_2},\ldots,\omega_{\alpha_m}\}$ and $\mathbf{\tilde{\omega}}_{\beta}=\{\omega_{\beta_1},\omega_{\beta_2},\ldots,\omega_{\beta_m}\}$, and $\mathcal{Z}$ denotes a suitable function space incorporating appropriate boundary conditions for the problem being solved. The auxiliary potential fields, $\omega_{\alpha_j}$ and $\omega_{\beta_j}$ for $j=1\ldots m$, introduced in the local reformulation of kernel energy are referred to as the \emph{kernel potentials}.

Using the reformulations in equations~\eqref{eqn:elReformulation} and ~\eqref{eqn:kerReformulation}, the extended interactions in electrostatics and the kernel kinetic energy functional can be recast into local variational problems in auxiliary potential fields---electrostatic potential and kernel potentials. Thus, using these local reformulations, the problem of evaluating the ground-state energy and electron density reduces to a local saddle point problem in the \emph{electronic fields} comprising of electron density, electrostatic potential and kernel potentials (cf. \citet{GaviniPRB2010,Motamarri-OFDFT2012}). Such a local reformulation in real-space is crucial for employing non-homogenous Dirichlet boundary conditions for studying the energetics of isolated defects in bulk, which is exploited in this work (elaborated subsequently in Section~\ref{sec:Results}).

\subsection{Finite-element discretization}

Plane wave basis has been the popular choice for the numerical discretization of the orbital-free DFT variational problem~\citep{Yan2,PROFESS2} owing to the ease of computing the extended electrostatic interactions and the kernel kinetic energy functional in Fourier-space using Fourier transforms. However, Fourier-space methods are limited to periodic systems, which poses a severe restriction in studying defects in materials. In particular, the geometry of an isolated dislocation is not periodic, and, thus, orbital-free DFT simulations on dislocations have either been limited to an artificial dipole/quadrapole arrangements of dislocations, or free-surfaces are introduced to study isolated dislocations~\citep{Carter2009,Shin2012}. While these studies have provided useful insights into the core-structure of dislocations, a quantification of the core-energies has been beyond reach. In order to overcome these limitations, recent efforts have also focused on developing real-space discretization techniques for the orbital-free DFT problem, which include finite difference discretization~\citep{Carlos,Phanish} and the finite-element discretization~\citep{Gavini1,Motamarri-OFDFT2012}. In particular, the finite-element discretization is attractive as it can handle complex geometries and boundary conditions, which is a crucial aspect that is exploited in this study. Further, recent investigations have demonstrated the accuracy and computational efficiency afforded by higher-order finite-element discretization for density functional theory calculations~\citep{Motamarri-OFDFT2012,Motamarri-KSDFT2013}, which, along with the good parallel scalability of finite-element discretizations, make the finite-element basis a desirable choice for real-space orbital-free DFT calculations. We refer to \citet{Motamarri-OFDFT2012} for a comprehensive discussion on the numerical analysis of the finite-element discretization of the orbital-free DFT problem, and the numerical aspects of the solution of the discrete orbital-free DFT problem.


\section{Results and Discussion}\label{sec:Results}

We now present our study of an edge dislocation in Aluminum using large-scale electronic-structure calculations. In particular, we unambiguously identify the size of the dislocation core, directly compute the dislocation core-energy, study the effect of macroscopic deformations on dislocation core-energy and core-structure, and investigate its implications on dislocation behavior. In this study, we employ orbital-free density functional theory calculations using the Wang-Govind-Carter (WGC) kinetic energy functional~\citep{Yan2} (second order Taylor expansion of the density dependent kernel, cf.~\citet{Yan2}), a local density approximation (LDA) for the exchange-correlation energy~\citep{Perdew}, and Goodwin-Needs-Heine pseudopotential~\citep{Goodwin}. While orbital-free DFT uses an approximation for the kinetic energy functional, it has been demonstrated that the WGC kinetic energy functional is in good agreement with Kohn-Sham DFT for a wide range of material properties in Aluminum, which include bulk and defect properties~\citep{Yan2,Ho,Carter-Al-Mg,Carter2009,Shin2012}. Further, orbital-free DFT is inherently linear-scaling in the number of electrons and enables the consideration of large system sizes, which is necessary, as will be demonstrated in this work, for an accurate study of the energetics of dislocations.

In the present work, we employ recently developed local real-space formulation of orbital-free DFT (cf. Section~\ref{sec:OFDFT} and ~\cite{Gavini1,GaviniPRB2010}) and a finite-element discretization of the formulation~\citep{Motamarri-OFDFT2012} to compute the energetics of an edge dislocation in Aluminum. In this real-space formulation, all the extended interactions, which include the electrostatic interactions and the kernels energies in the WGC kinetic energy functional, are reformulated into local variational problems in auxiliary potential fields that include the electrostatic potential and kernel potentials. The finite-element basis employed in the present work enables consideration of complex geometries and general boundary conditions (Dirichlet, periodic and semi-periodic), which is another crucial aspect that enables an accurate quantification of the energetics of isolated dislocations. The accuracy of the local real-space formulation of orbital-free DFT and the convergence of the finite-element discretization of the formulation have been demonstrated for both bulk and defect properties in Aluminum~\citep{GaviniPRB2010,Motamarri-OFDFT2012}. In the present study, we use quadratic hexahedral finite-elements in all our calculations, where the basis functions correspond to a tensor product of quadratic polynomials in each dimension. The finite-element discretization and other numerical parameters---quadrature rules and stopping tolerances on iterative solvers---are chosen such that the errors in the computed dislocation energies are less than $0.005$ $eV$. In all our calculations, the atomic relaxations are performed till the maximum force on any atom is less than $0.01$ $eV/\mbox{\AA}$.

\subsection{Dislocation core-size and core-energy} In order to compute the dislocation core-energy, we begin by first estimating the dislocation core-size. Continuum theories estimate the dislocation core-sizes based on displacement fields, and prior studies suggest the dislocation core-size to be $\sim$ $1-3$ $|\mathbf{b}|$~\citep{Hirth,Peierls1940,Kioussis2007,Weinberger2008}, where $\mathbf{b}$ denotes the Burgers vector. While the displacement and strain fields may be well described by elastic formulations outside of a $1-3$ $|\mathbf{b}|$ core region, the perturbations in electronic-structure due to the dislocation may be present on a larger region. The energetic contributions from these electronic-structure effects can not be captured in continuum elasticity theories or atomistic calculations using empirical potentials. Thus, in the present work, we consider the dislocation-core to be the region where electronic-structure effects and their contribution to the defect energy are significant.

We consider a perfect edge dislocation in face-centered-cubic Aluminum to determine the dislocation core-size. We align our coordinate system, $X-Y-Z$ axes (or equivalently $1-2-3$), along $[1\,1\,0]-[1\,\overline{1}\,1]-[1\,\overline{1}\,\overline{2}]$ crystallographic directions, respectively. We begin by considering a perfect crystal of size $2R\sqrt{2}a_0\times 2R\sqrt{3}a_0 \times0.5\sqrt{6}a_0$ where $a_0$ denotes the lattice parameter and $R$ is an integer-valued scaling factor used to consider a sequence of increasing simulation domain sizes. A perfect edge dislocation with Burgers vector $\mathbf{b}=\frac{a_0}{2}[110]$ is introduced at the center of the simulation domain by removing two consecutive half-planes normal to $[110]$ and applying the continuum displacement fields of an edge dislocation to the positions of atoms. We note that this approach of creating the edge dislocation is devoid of the asymmetry that can otherwise arise by applying the Volterra solution in the most obvious manner by creating slip (cf.~\citet{Gehlen1978}). In this work, the displacement fields from isotropic elasticity~\citep{Hirth} are employed as anisotropic effects are not strong in Aluminum. Upon the application of displacement fields, the geometry of the computational domain is no longer cuboidal, and, thus, the use of a finite-element basis which can accommodate complex geometries is crucial to the present study.

In order to simulate a perfect edge dislocation, we hold the positions of the atoms fixed and compute the electronic-structure using orbital-free DFT. We employ Dirichlet boundary conditions on the electronic fields---comprising of electron density, electrostatic potential and kernel potentials---in the $X$ and $Y$ directions and use periodic boundary conditions along the $Z$ direction. The Dirichlet boundary conditions are determined under the Cauchy-Born approximation, where the values of electronic fields on the Dirichlet boundary are computed from periodic unit-cells undergoing the deformations produced by the edge dislocation. The electronic-structure, thus computed, represents an isolated edge dislocation in bulk with the electronic-structure perturbations from the edge dislocation vanishing on the Dirichlet boundary. The local real-space formulation of orbital-free DFT and the finite-element basis are key to employing these boundary conditions, which are not possible to realize in Fourier-space based formulations.

\begin{figure}[htbp]%
\centering%
\scalebox{0.33}{\includegraphics{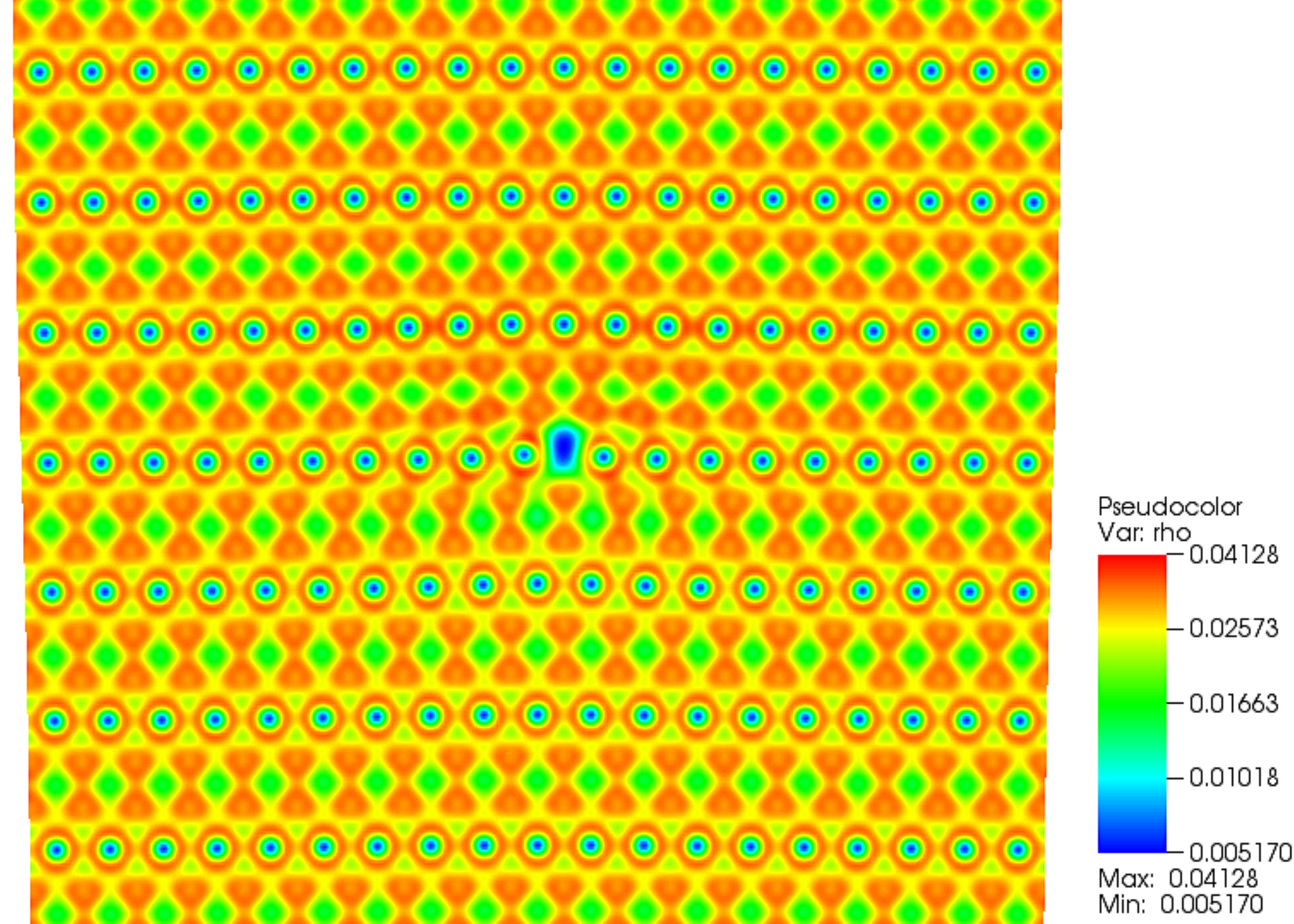}}%
\caption{\label{fig:contours} Electron density contours of a perfect edge dislocation in Aluminum.}%
\end{figure}

\begin{figure}[htbp]
\centering
\includegraphics[width=0.65\textwidth]{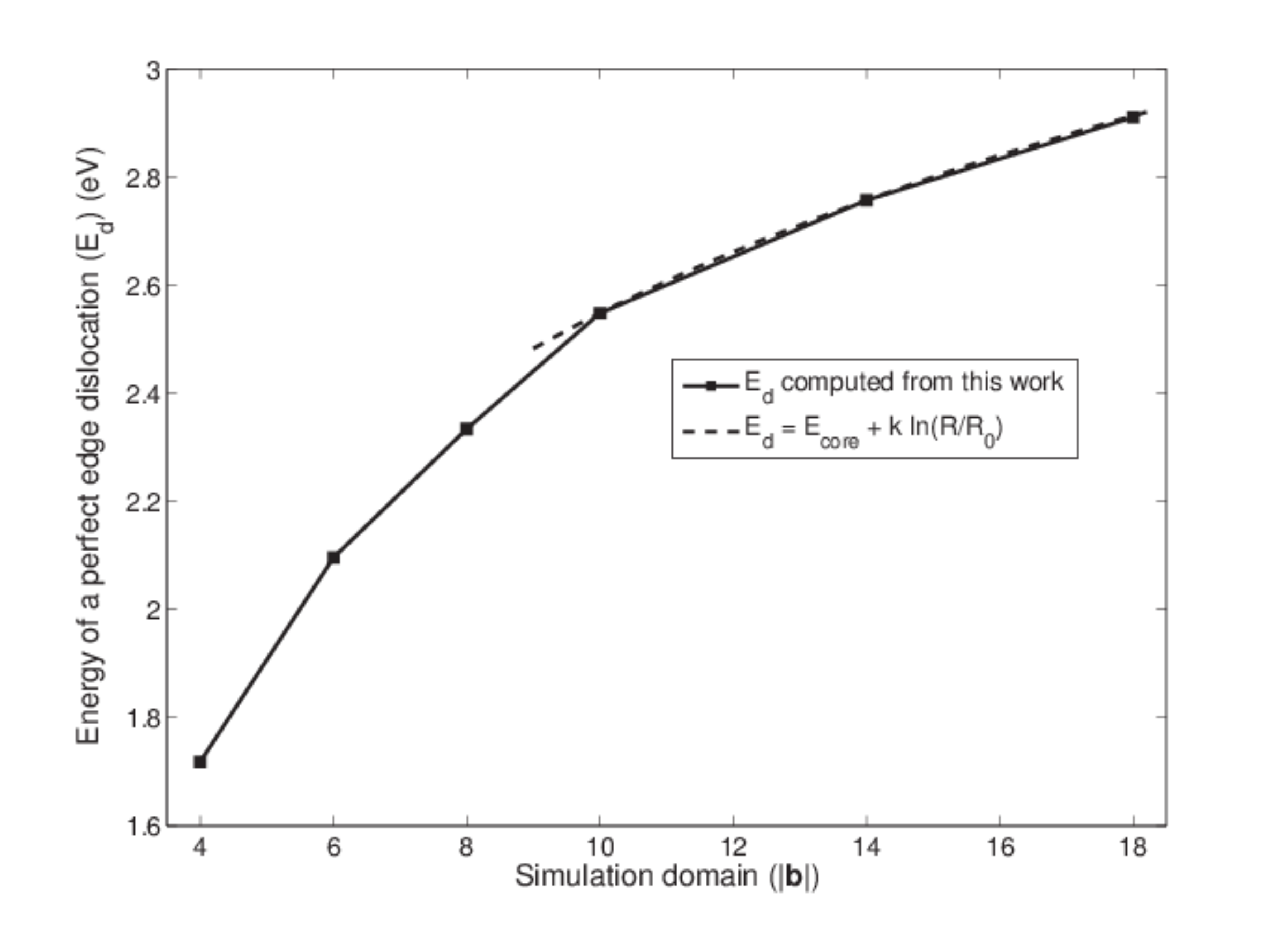}
\caption{Energy of a perfect edge dislocation as a function of simulation domain size. The dashed-line demonstrates the asymptotic logarithmic divergence of the computed dislocation energies, as expected from continuum estimates, beyond simulation domains of $10|\mathbf{b}|$ (corresponding to a domain of $10\sqrt{2}a_{0}\times 10\sqrt{3}a_0 \times 0.5\sqrt{6}a_0$ along $[110]-[1\overline{1}1]-[1\overline{1}\overline{2}]$).}
\label{fig:cellSize}
\end{figure}

We computed the electronic-structure and ground-state energy of the perfect edge dislocation for varying simulation domains with $R=2,3,4,5,7,9$. Figure~\ref{fig:contours} shows the contours of the electron density for $R=5$. We note that a scaling factor $R$ corresponds to a domain-size where the distance from dislocation line to the boundary along $[110]$ is $2R|\mathbf{b}|$. The dislocation energy ($E_d$) for these various simulation domains is computed as
\begin{equation}\label{eqn:dislocEnergy}
E_{d} (N,V)=E_{disloc}(N,V)-E_0(N,V)\,,
\end{equation}
where $E_{disloc}(N,V)$ denotes the energy of the $N$ atom system comprising of the dislocation and occupying a volume $V$, and $E_0(N,V)$ denotes energy of a perfect crystal containing the same number of atoms and occupying the same volume. We first computed the dislocation energy at equilibrium volume (i.e. $V=N\frac{a_0^3}{4}$) for the various domain-sizes considered in this study, and the results are presented in Table~\ref{tab:elasticElec}. The computed dislocation energy increases with increasing domain-size, and has an asymptotic logarithmic divergence as expected from continuum theories. In order to understand the extent of electronic relaxations, we consider the change in dislocation energy by increasing the domain-size---for instance, from a domain-size of $2R_1|\mathbf{b}|$ to $2R_2|\mathbf{b}|$---and denote this change by $\Delta E_{d}$. This change in the energy has two contributions: (i) the increase in the elastic energy due to the increase in the domain-size, which we denote by $\Delta E_{d}^{elas}$; (ii) electronic contribution from perturbations in the electronic-structure, which we denote by $\Delta E_{d}^{elec}$. As the elastic energy contribution to the dislocation energy is due to the elastic deformation fields produced by the dislocation, we estimate $\Delta E_{d}^{elas}$ by integrating the elastic energy density in the region of interest. The elastic energy density at any point, in turn, is computed from an orbital-free DFT calculation on a unit-cell undergoing the macroscopic deformation produced by the edge dislocation at that point. We verified that the discretization errors in the computation of $\Delta E_{d}^{elas}$ are of the order of $0.001~eV$. Upon computing $\Delta E_{d}^{elas}$, we infer $\Delta E_{d}^{elec}$ from $\Delta E_{d}$ ($\Delta E_{d}^{elec}=\Delta E_{d} -\Delta E_{d}^{elas}$). The computed $\Delta E_{d}^{elas}$ and $\Delta E_{d}^{elec}$ are reported in Table~\ref{tab:elasticElec}. It is interesting to note from the results that $\Delta E_{d}^{elec}$ is comparable to $\Delta E_{d}^{elas}$ up to a domain-size of $10|\mathbf{b}|$, suggesting that the electronic-structure perturbations are significant up to distances as far as $10|\mathbf{b}|$ from the dislocation line. This result is in strong contrast to conventional estimates of core-sizes to be $1-3~|\mathbf{b}|$. Further, we note that a core-size of $10|\mathbf{b}|$ contains over $1000$ atoms, which is much larger than the computational cells employed in most prior electronic-structure studies of dislocations, and underscores the need to consider sufficiently large simulation domains to accurately compute the energetics of dislocations. In the remainder of this work, we consider $10|\mathbf{b}|$ to be the core-size of edge dislocation, and the dislocation energy corresponding to this core-size as the dislocation core-energy. For a perfect edge dislocation, the computed core-energy is $2.548~eV$, or, equivalently, the core-energy per unit length of dislocation line is $0.515~eV/\mbox{\AA}$. A plot of the dislocation energies ($E_d$) reported in Table~\ref{tab:elasticElec} as a function of domain-size is provided in figure~\ref{fig:cellSize}, which shows the expected asymptotic logarithmic divergence of the dislocation energy.

\begin{table}[htbp]
\caption{\label{tab:elasticElec}Computed dislocation energy of perfect edge dislocation in Aluminum for varying domain-sizes, where $N$ denotes the number of atoms in the simulation domain. $\Delta E_{d}$ denotes the change in the dislocation energy from the previous domain-size, and $\Delta E_{d}^{elas}$ and $\Delta E_{d}^{elec}$ denote the elastic and electronic contributions to $\Delta E_{d}$.}
\centering 
\scalebox{1.0}{
\begin{tabular} {c  c  c  c  c c}
\hline
Simulation & $N$ & $E_{d} $ & $\Delta E_{d}$ & $\Delta E_{d}^{elas}$ & $\Delta E_{d}^{elec}$ \\
    domain      &  (atoms)    &  (eV)         &   (eV)   &      (eV)     &      (eV)\\ \hline
4$\left|\mathbf{b}\right|$ & 179 & 1.718 &    -    &    -     &    -     \\
6$\left|\mathbf{b}\right|$ & 413 & 2.096 & 0.378 & 0.230 & 0.148  \\
8$\left|\mathbf{b}\right|$ & 743 & 2.334 & 0.238 & 0.164 & 0.074  \\
10$\left|\mathbf{b}\right|$ & 1169 & 2.548 & 0.214 & 0.118 & 0.096  \\
14$\left|\mathbf{b}\right|$ & 2309 & 2.757 & 0.209 & 0.187 & 0.022  \\
18$\left|\mathbf{b}\right|$ & 3833 & 2.91 & 0.153 & 0.156 & -0.003  \\\hline
\end{tabular}
}
\end{table}

\begin{figure}[htbp]%
\centering
\subfigure{
\scalebox{0.5}{\includegraphics{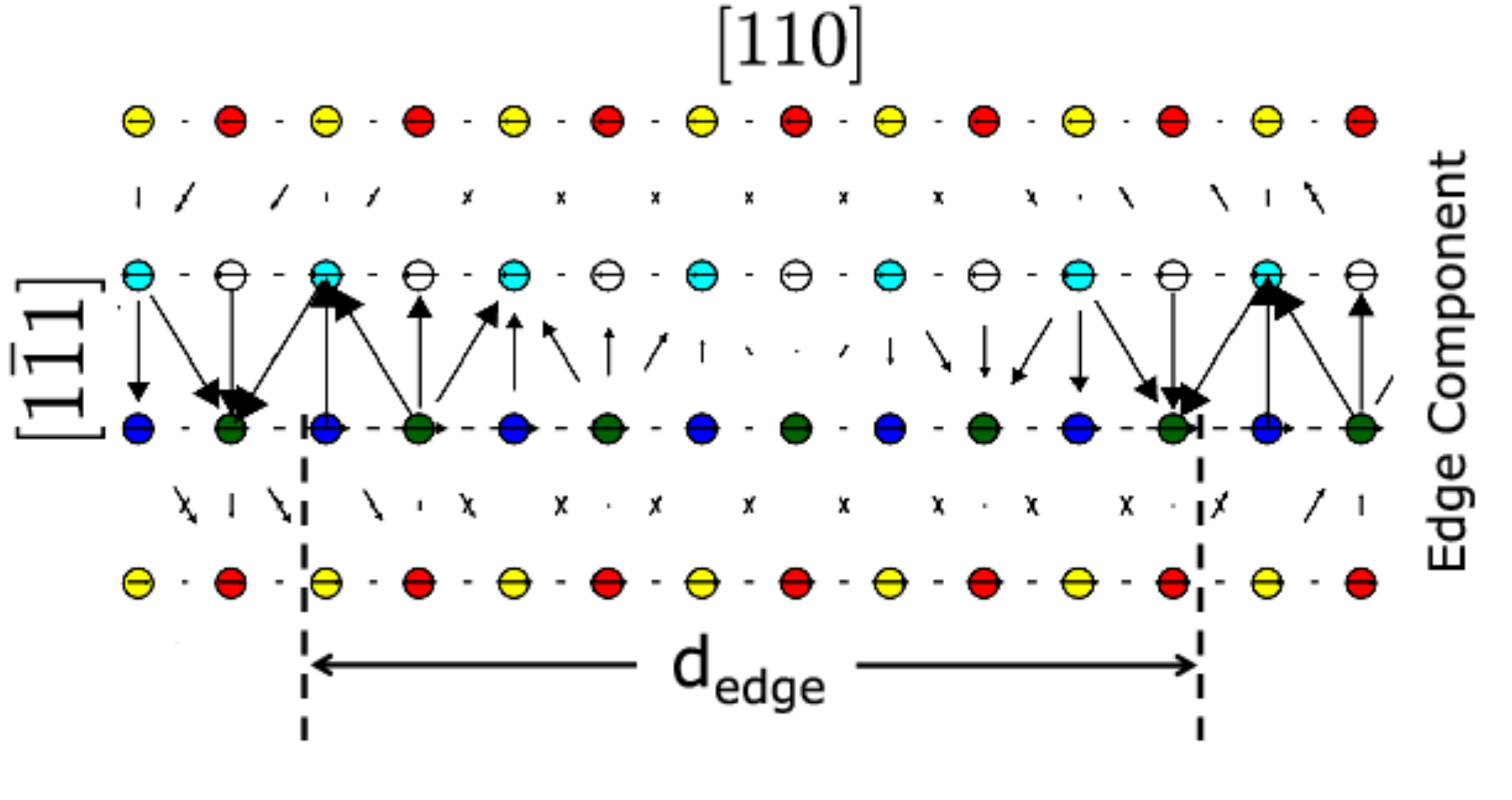}}
\label{fig:DD_edge}
}
\subfigure{
\scalebox{0.5}{\includegraphics{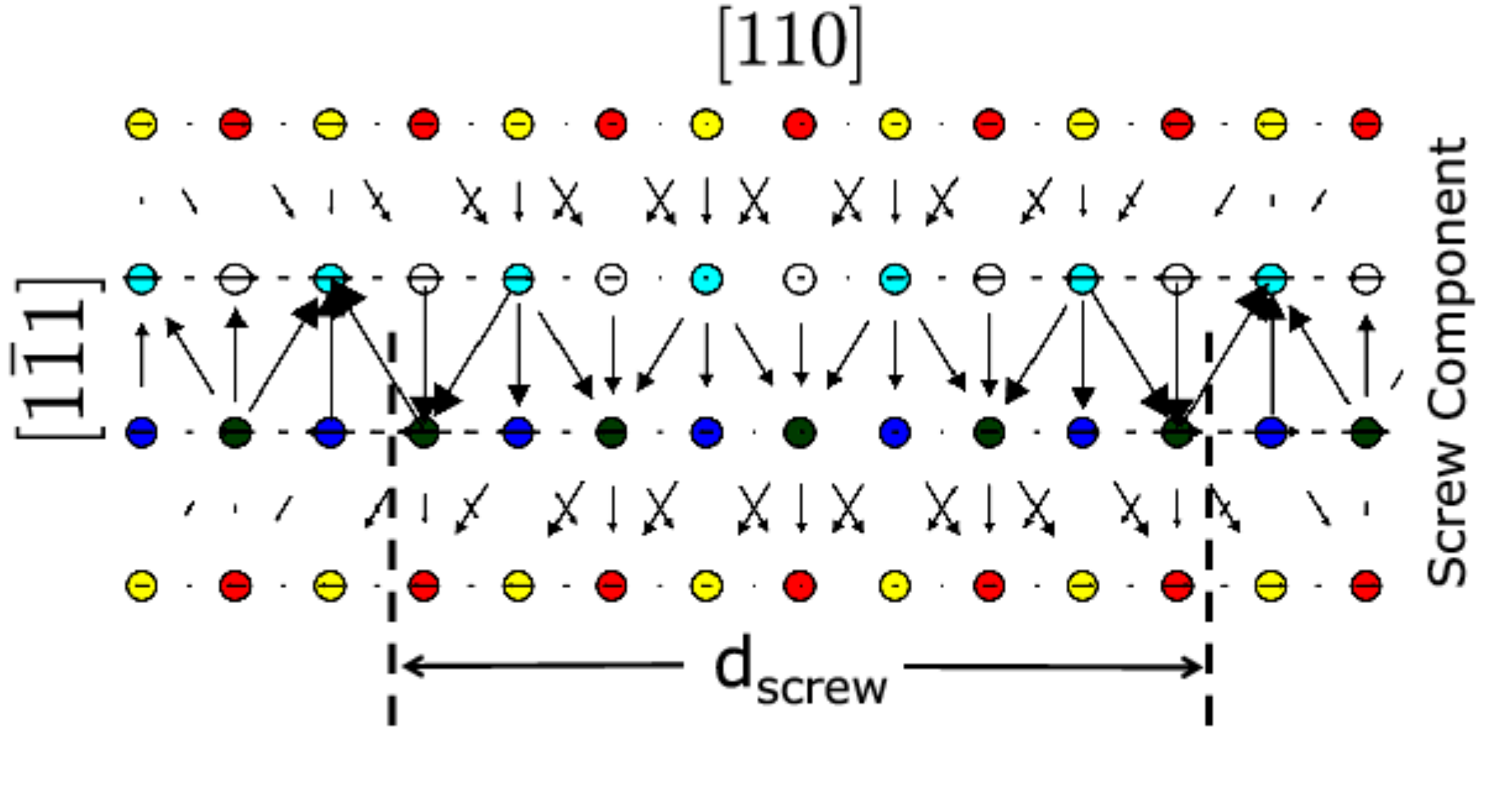}}
\label{fig:DD_screw}
}
\caption{\label{fig:DD} Differential displacement plots of the edge and screw components of Shockley partials. The dotted lines represent the location of the partials.}%
\end{figure}

As a next step in our study, we allowed for internal atomic relaxations in the simulation domain corresponding to $10|\mathbf{b}|$ by holding the positions of atoms fixed on the boundary. Upon atomic relaxations, the perfect edge dislocation split into Shockley partials. Figure~\ref{fig:DD} shows the edge and screw components of the differential displacements~\citep{Vitek1970} and indicates the location of the Shockley partials. The partial separation distance computed from the edge-component differential displacement plot is $12.84$~\AA ~($4.5|\mathbf{b}|$), and that computed from the screw-component differential displacement plot is $11.39$~\AA ~($4|\mathbf{b}|$). This is in good agreement with other DFT studies on an edge dislocation in Aluminum which have reported partial separation distances between $9.5-13.7$~\AA~\citep{Trinkle2008,Carter2009}. The core-energy of Shockley partials is computed to be $1.983~eV$, or, equivalently, the core-energy per unit length of dislocation line is $0.401~eV/\mbox{\AA}$. The core-energy of the relaxed Shockley partials, as expected, is significantly lower than that of the perfect edge dislocation. In the remainder of this paper, all the reported core-energies correspond to the relaxed Shockley partials.

We note that holding the positions of atoms fixed on the boundary, prescribed by the displacement fields of a perfect edge dislocation, can result in finite cell-size effects that influence the core-structure of the Shockley partials and the core-energy. In order to investigate the influence of this possible finite cell-size effect, we considered simulation domains corresponding to $12|\mathbf{b}|$ and $14|\mathbf{b}|$ and computed the core-structure in these simulation domains, and compared it with the core-structure obtained from the $10|\mathbf{b}|$ simulation domain. The partial separation distance of both the edge- and screw-components did not change by increasing the simulation domain. Further, we computed the difference in the positions of atoms (in the region corresponding to $10|\mathbf{b}|$ simulation domain) obtained from the $14|\mathbf{b}|$ and $10|\mathbf{b}|$ simulation domains. The maximum difference in the positions of atoms is $\sim 0.01 \mbox{\AA}$, which is close to the tolerance associated with force relaxations. These results suggest that the finite cell-size effects arising from the fixed spatial boundary conditions on atomic positions are not important in simulation domains of size $10|\mathbf{b}|$ and beyond.

While the computed partial separation distance is in good agreement with prior electronic-structure studies, the partial separation distances are over-predicted in DFT compared to experimental investigations. The partial separation distance of the edge dislocation in Aluminum is estimated to be $\sim 8 \mbox{\AA}$ using weak-field TEM~\citep{Hollerbauer}. One plausible cause of this deficiency in DFT calculations can be the use of pseudopotentials, especially since the stacking fault energy is  sensitive to the choice of the pseudopotential~\citep{Trinkle2008,Carter2008}. While pseudopotential DFT calculations have been shown to be accurate for the prediction of bulk properties in solids, their transferability to accurately predict defect properties is yet to been ascertained. To this end, there is a growing need for the development of efficient large-scale real-space methods for all-electron Kohn-Sham DFT calculations, and we refer to~\citet{Motamarri-KSDFT2013, MotamarriPRB2014} for recent efforts in this direction.

\subsection{Effect of macroscopic deformations}
Most continuum and mesoscale models describing deformation and failure mechanisms in crystalline solids account for the role of defects through the elastic interactions between various defects, as well as, elastic interactions between defects and macroscopically applied loads. However, recent electronic-structure studies on point defects~\citep{GaviniPRL2008,GaviniPRSA2009,Iyer2014} suggest that the defect-core can play a significant role in governing the overall energetics of these defects, and that there is a strong influence of macroscopic deformations on the electronic-structure of the defect-core and its energetics. Further, a recent electronic-structure study also reported a sizeable effect of pressure on the dislocation-core properties in semiconductor materials~\citep{Pizzagalli2009}.

Thus, as a next step in our study, we investigate the effect of macroscopic deformations on the core-energy and core-structure of an edge dislocation in Aluminum. We begin with a perfect edge dislocation in $10|\mathbf{b}|$ simulation domain, which corresponds to the core-size of the edge dislocation. We subsequently apply an affine deformation on the simulation domain corresponding to a macroscopic strain $\varepsilon_{ij}$, and, by holding the positions on the boundary fixed, compute the electronic-structure as well as the atomic structure of the relaxed Shockley partials. In these simulations, as discussed earlier, Dirichlet boundary conditions are applied on the electronic fields in the $X$ and $Y$ directions, where the boundary values are computed using the Cauchy-Born hypothesis. We note that the applied macroscopic strain manifests into the boundary conditions on the positions of atoms, as well as, the values of the electronic fields on the Dirichlet boundary. Following equation~\eqref{eqn:dislocEnergy}, the dislocation energy of this $10|\mathbf{b}|$ simulation domain (dislocation core-energy) as function of macroscopic strain is computed as
\begin{equation}
E_d(\epsilon_{ij})=E_{disloc}(\epsilon_{ij})-E_{0}(\epsilon_{ij})\,.
\end{equation}
In the above expression, $E_{disloc}(\epsilon_{ij})$ denotes the ground-state energy (after internal atomic relaxations) of the $10|\mathbf{b}|$ simulation domain, which contains the dislocation, under an affine deformation corresponding to macroscopic strain $\epsilon_{ij}$, and $E_{0}(\epsilon_{ij})$ denotes the energy of a perfect crystal under the same affine deformation, containing the same number of atoms and occupying the same volume.

\begin{figure}[htbp]%
\centering%
\scalebox{1.13}{\includegraphics[trim=3.5cm 0cm 0cm 0cm ,clip=true,width=\textwidth]{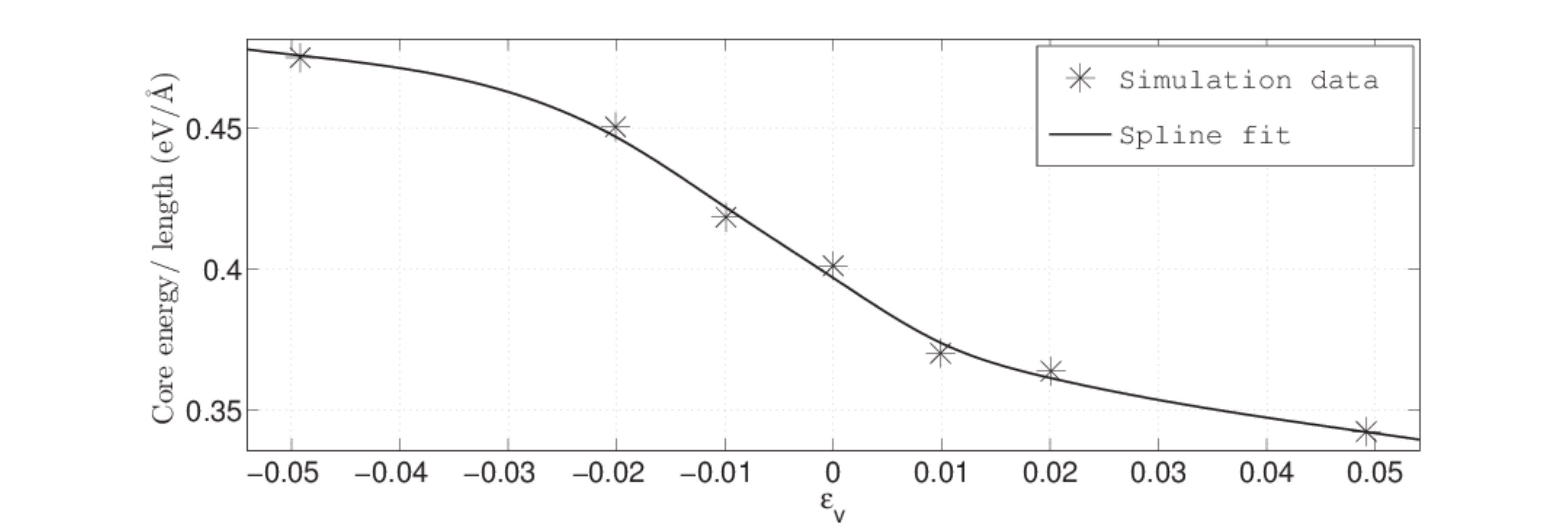}}
\caption{\label{fig:volumetric_strain} Core-energy per unit length of dislocation line of relaxed Shockley partials as a function of macroscopic volumetric strain.}%
\end{figure}

We begin our investigation by studying the effect of macroscopic volumetric strain ($\varepsilon_{v}$), corresponding to equi-triaxial strain, on the core-energy of Shockley partials and the core-structure. In this study, we consider volumetric strains of $-5\%$, $-2\%$, $-1\%$, $1\%$, $2\%$ and $5\%$. The computed core-energy demonstrated a strong dependence on volumetric strain, and is shown in figure~\ref{fig:volumetric_strain}. The core-energy (per unit length of dislocation line) changed from $0.48~eV/\mbox{\AA}$ at $-5\%$ volumetric strain to $0.34~eV/\mbox{\AA}$ at $5\%$ volumetric strain, and this change corresponds to a significant fraction ($\sim 0.3$) of the core-energy at equilibrium. This finding is in sharp contrast to the assumptions in continuum formulations that ignore the core-energy in determining the energetics of dislocations and their interactions. The core-structure, analyzed using the differential displacement (DD) plots, is also found to be marginally influenced by the applied macroscopic volumetric strain. In particular, for the range of volumetric strains considered, the partial separation in the edge-component DD plots is in the range of $4.5-5~|\mathbf{b}|$, and the partial separation in the screw-component DD plot is in the range of $3.5-4~|\mathbf{b}|$.

\begin{figure}[htbp]%
\centering%
\subfigure[ ]{
\scalebox{1.12}{\includegraphics[trim=2.5cm 0cm 0cm 0cm ,clip=true,width=\textwidth]{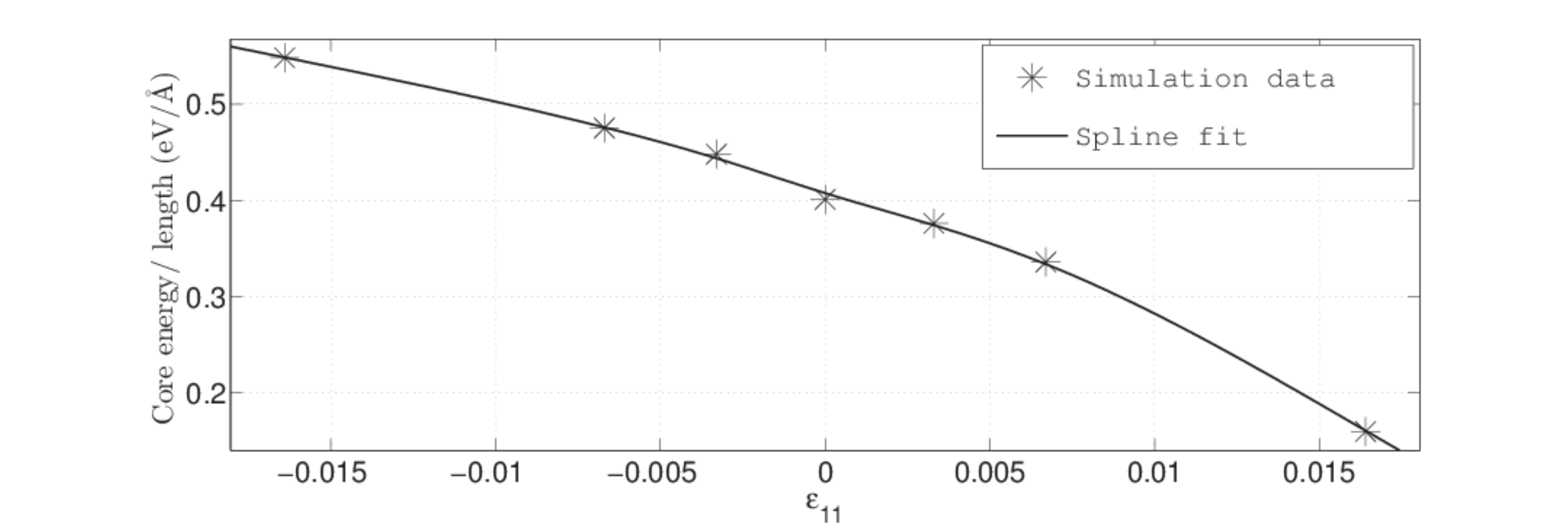}}
\label{fig:uniaxial1}}
\subfigure[ ]{
\scalebox{1.12}{\includegraphics[trim=2.5cm 0cm 0cm 0cm ,clip=true,width=\textwidth]{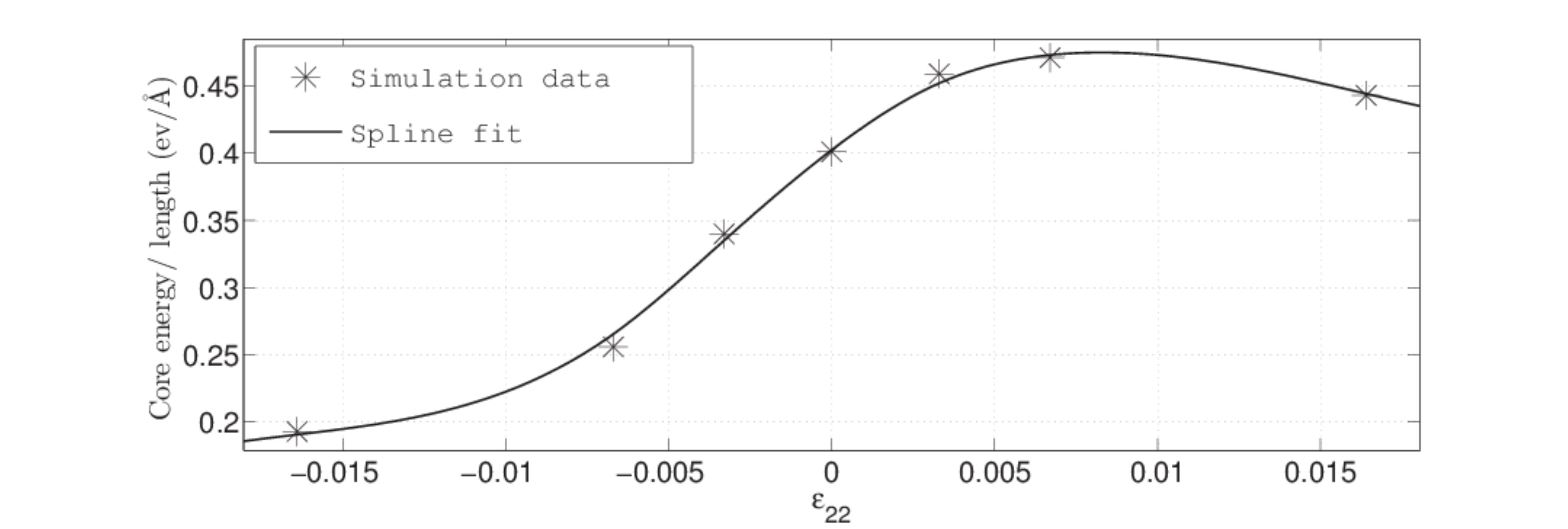}}
\label{fig:uniaxial2}}
\subfigure[ ]{
\scalebox{1.12}{\includegraphics[trim=2.5cm 0cm 0cm 0cm ,clip=true,width=\textwidth]{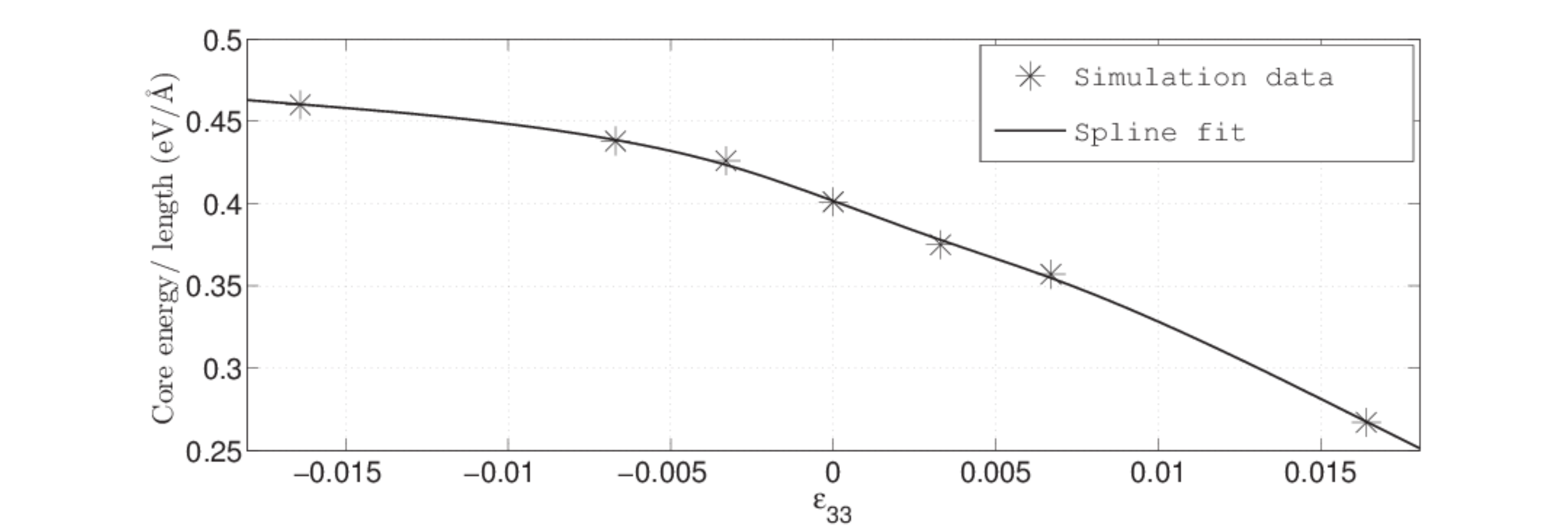}}
\label{fig:uniaxial3}}
\caption{\label{fig:uniaxial_strain} Core-energy per unit length of dislocation line of relaxed Shockley partials as a function of uniaxial strains: (a) $\varepsilon_{11}$ (along $[1\,1\,0]$); (b) $\varepsilon_{22}$ (along $[1\,\overline{1}\,1]$); (c) $\varepsilon_{33}$ (along $[1\,\overline{1}\,\overline{2}]$).}%
\end{figure}

We next study the effect of uniaxial strains along the coordinate directions, $[1\,1\,0]-[1\,\overline{1}\,1]-[1\,\overline{1}\,\overline{2}]$, on the dislocation core-energy and core-structure. We consider uniaxial strains of $-1.64\%$, $-0.66\%$, $-0.33\%$, $0.33\%$, $0.66\%$ and $1.64\%$ along each of the coordinate directions. Figures~\ref{fig:uniaxial1},~\ref{fig:uniaxial2}, and \ref{fig:uniaxial3} show the core-energy dependence on $\varepsilon_{11}$ (uniaxial strain along $[1\,1\,0]$), $\varepsilon_{22}$ (uniaxial strain along $[1\,\overline{1}\,1]$) and $\varepsilon_{33}$ (uniaxial strain along $[1\,\overline{1}\,\overline{2}]$), respectively. As in the case of volumetric strains, the core-energy monotonically decreases from compressive strains to tensile strains (for the range of strains considered) for uniaxial strains $\varepsilon_{11}$ (along the Burgers vector) and $\varepsilon_{33}$ (along the dislocation line). However, the core-energy dependence is non-monotonic and non-linear for $\varepsilon_{22}$ uniaxial strain (along the normal to the slip plane). These results show that the core-energy of an edge dislocation can have a complex and non-linear dependence on macroscopic strains. As in the case of volumetric strains, the core-structure is found to be marginally sensitive to uniaxial strains. For the range of uniaxial strains considered in this study, the partial separation in the edge-component DD plots is in the range of $4.5-5~|\mathbf{b}|$ and the partial separation in the screw-component DD plots is in the range of $4-4.5~|\mathbf{b}|$. The fact that the core-energy shows a strong dependence on the volumetric and uniaxial strains, while the core-structure is only marginally influenced by these strains, suggests that the electronic-structure at the core can play a dominant role in governing the energetics of dislocations and that the core-structure alone may not completely characterize the defect-core.

\begin{figure}[htbp]%
\centering%
\subfigure[ ]{
\scalebox{1.18}{\includegraphics[trim=2.0cm 0.5cm 0cm 0cm ,clip=true,width=\textwidth]{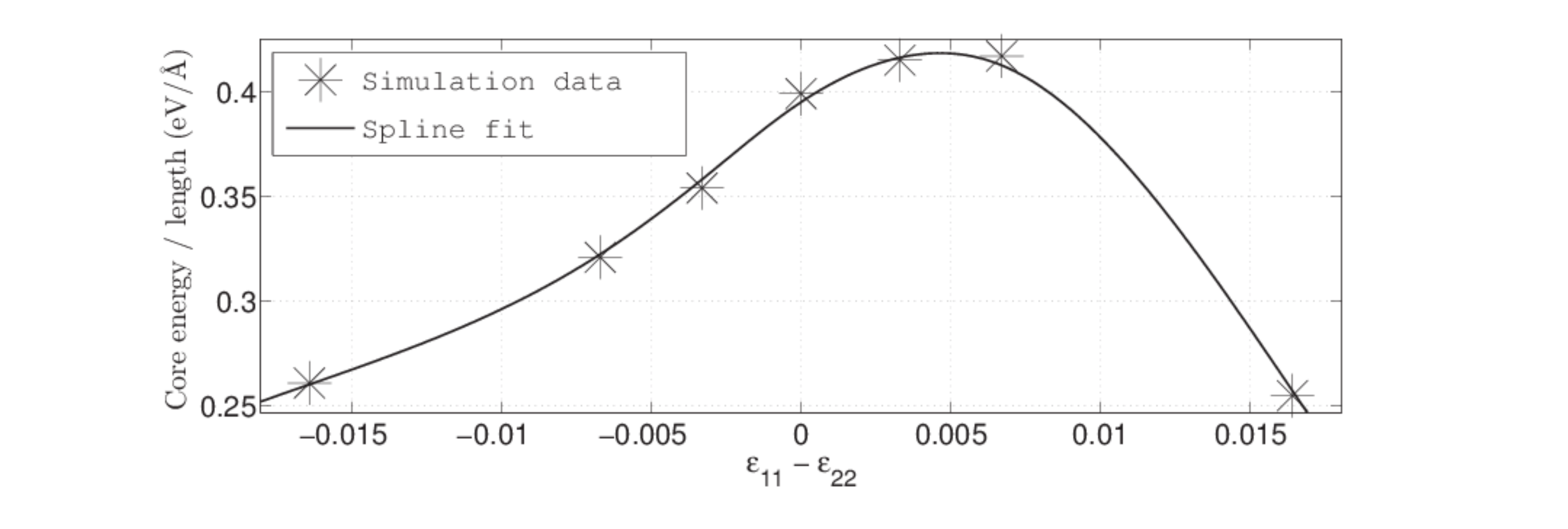}}
\label{fig:biaxial12}}
\subfigure[ ]{
\scalebox{1.1}{\includegraphics[trim=2.5cm 0cm 0cm 0cm ,clip=true,width=\textwidth]{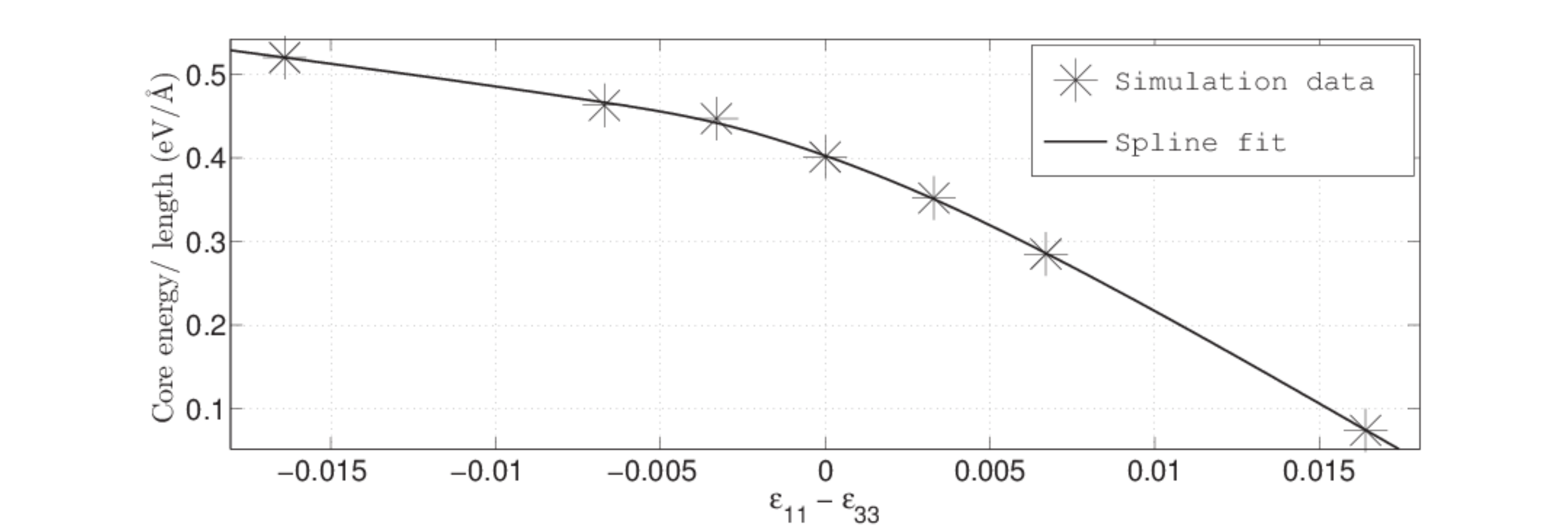}}
\label{fig:biaxial13}}
\subfigure[ ]{
\scalebox{1.1}{\includegraphics[trim=2.5cm 0cm 0cm 0cm ,clip=true,width=\textwidth]{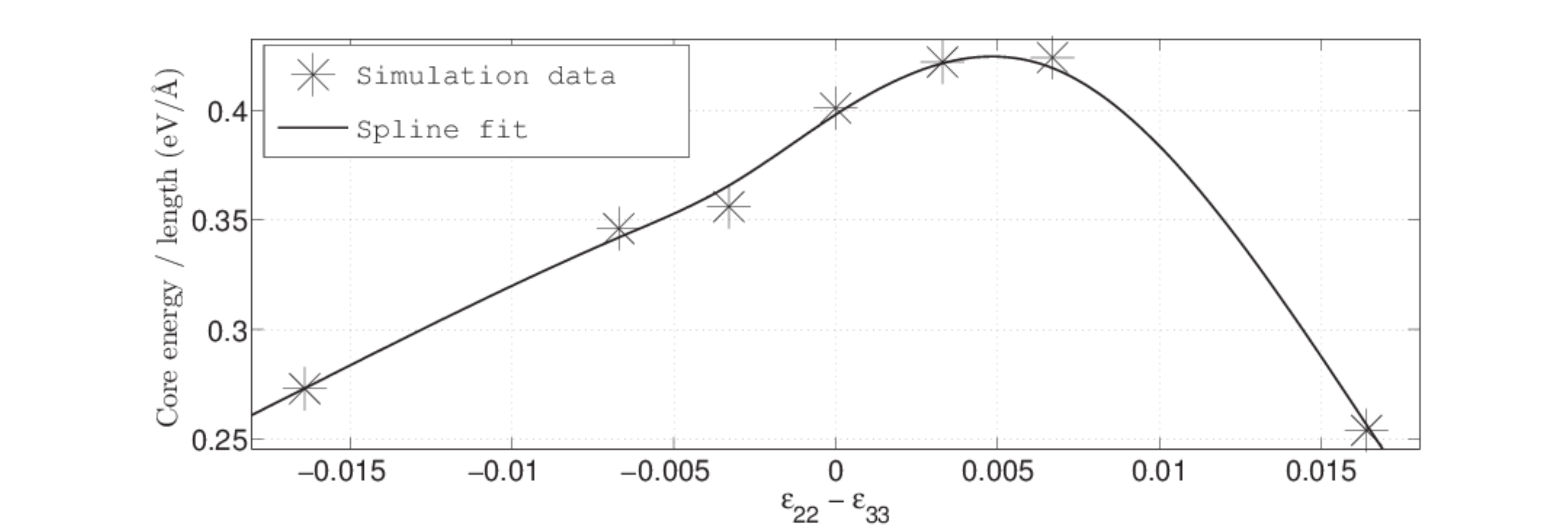}}
\label{fig:biaxial23}}
\caption{\label{fig:biaxial_strain} Core-energy per unit length of dislocation line of relaxed Shockley partials as a function of equi-biaxial strains: (a) $\varepsilon_{11}-\varepsilon_{22}$; (b) $\varepsilon_{11}-\varepsilon_{33}$; (c) $\varepsilon_{22}-\varepsilon_{33}$.}%
\end{figure}

We next investigate the effect of biaxial strains on the edge dislocation. To this end, we consider equi-biaxial states of strain along the coordinate directions, and computed the core-energies and core-structures under these applied macroscopic strains. Figures~\ref{fig:biaxial12}, ~\ref{fig:biaxial13}, ~\ref{fig:biaxial23} show the dependence of core-energy on equi-biaxial strains $\varepsilon_{11}-\varepsilon_{22}$, $\varepsilon_{11}-\varepsilon_{33}$ and $\varepsilon_{22}-\varepsilon_{33}$, respectively. As in the case of uniaxial strains, the core-energy dependence on biaxial strains is found to be non-monotonic when there is macroscopic strain along $[1\,\overline{1}\,1]$, the direction normal to the slip plane (cf. figures~\ref{fig:biaxial12}, ~\ref{fig:biaxial23}). These results again underscore the highly non-linear nature of core-energy dependence on macroscopic strains. The biaxial strains also influenced the core-structure, and to a greater extent than uniaxial strains. For the range of uniaxial strains considered in this study, the partial separation in the edge-component DD plots is in the range of $4.5-5.5~|\mathbf{b}|$ and the partial separation in the screw-component DD plots is in the range of $3.5-4.5~|\mathbf{b}|$.

\begin{figure}[htbp]%
\centering%
\subfigure[ ]{
\scalebox{1.1}{\includegraphics[trim=2.5cm 0cm 0cm 0cm ,clip=true,width=\textwidth]{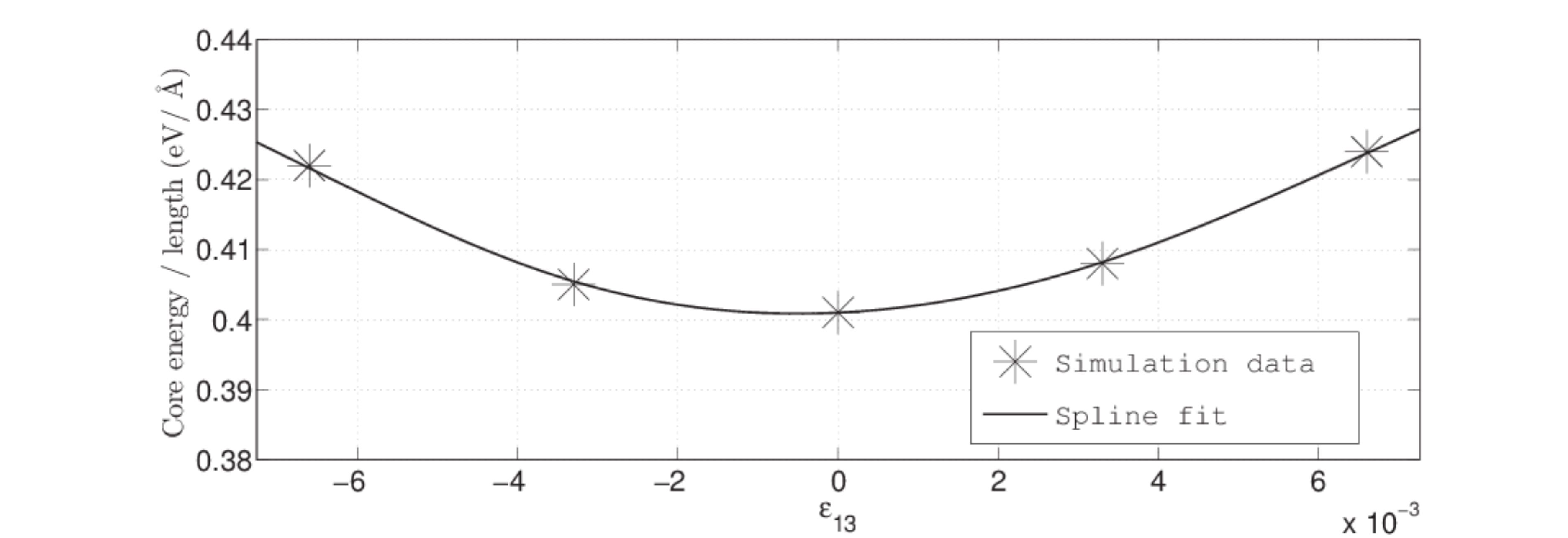}}
\label{fig:shear13}}
\subfigure[ ]{
\scalebox{1.1}{\includegraphics[trim=2.5cm 0cm 0cm 0cm ,clip=true,width=\textwidth]{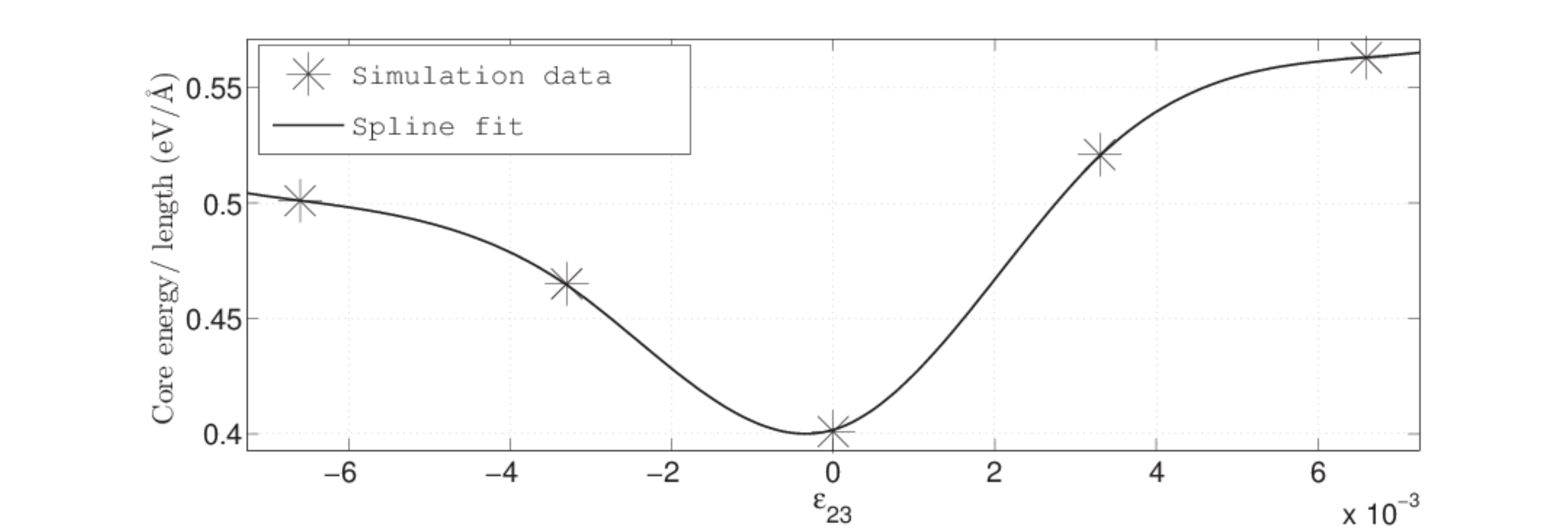}}
\label{fig:shear23}}
\caption{\label{fig:shear_strain} Core-energy per unit length of dislocation line of relaxed Shockley partials as a function of shear strains: (a) $\varepsilon_{13}$; (b) $\varepsilon_{23}$.}%
\end{figure}

\begin{figure}[htbp]%
\centering
\subfigure[]{
\scalebox{0.55}{\includegraphics[trim=0cm 0cm 0cm 0cm]{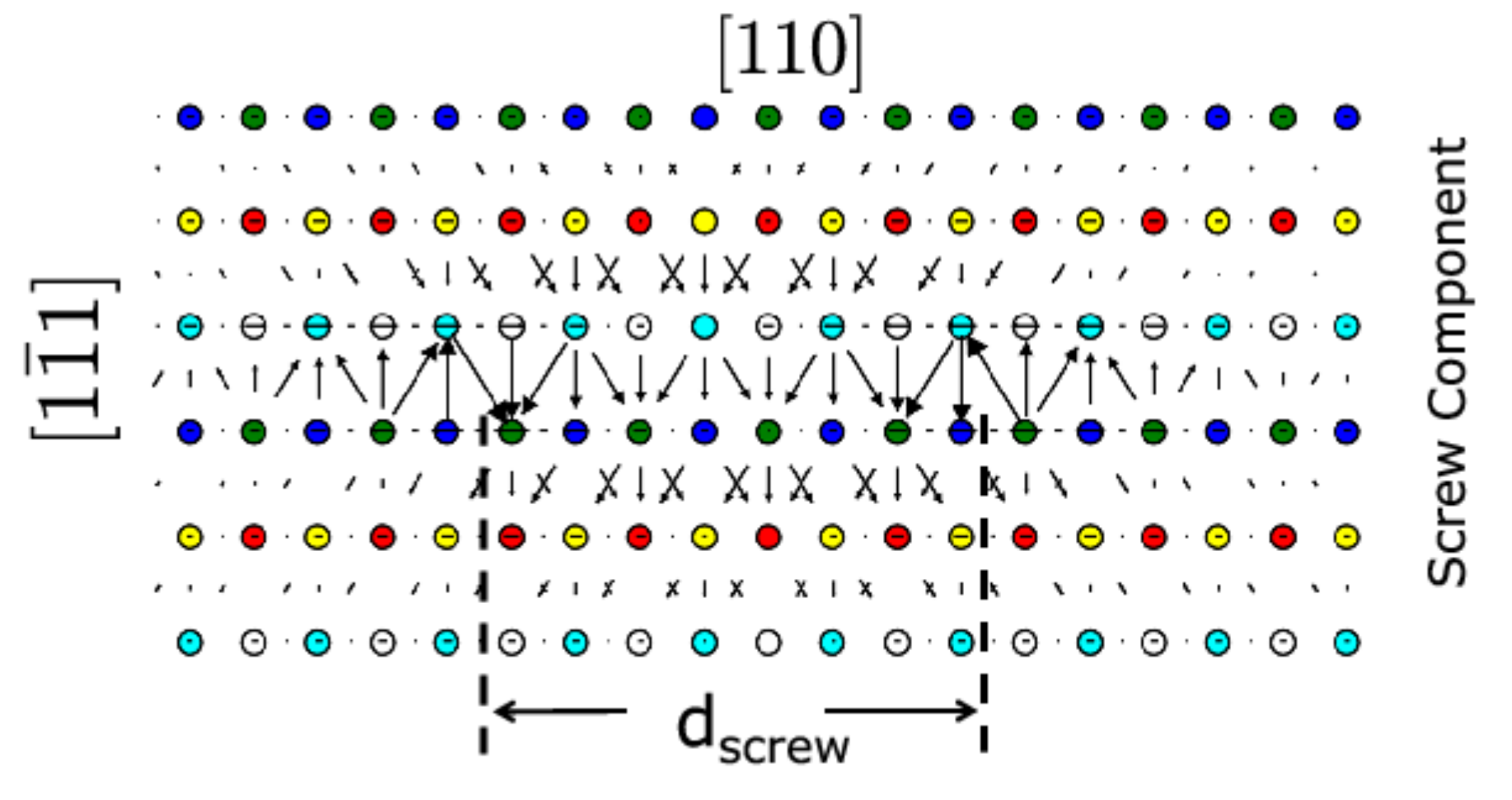}}
\label{fig:DD_screw_e23_m066}
}
\subfigure[]{
\scalebox{0.55}{\includegraphics[trim=0cm 0cm 0cm -1cm]{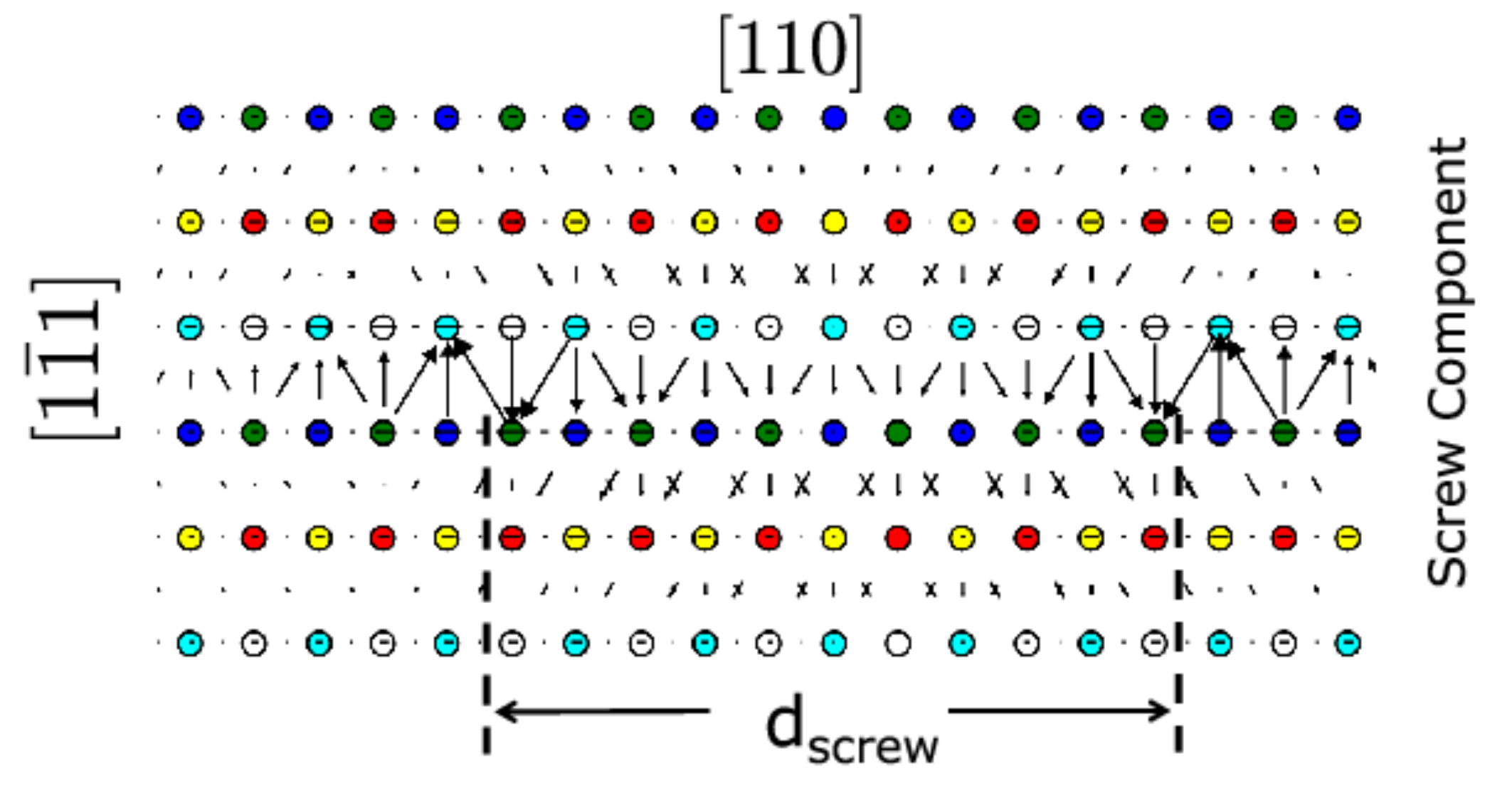}}
\label{fig:DD_screw_e23_p066}
}
\caption{\label{fig:DD_screw_e23} Differential displacement plots of the screw components of Shockley partials: (a) $\epsilon_{23} =-0.66$\,; (b) $\epsilon_{23} =0.66$. The dotted lines represent the location of the partials.}%
\end{figure}

We next consider shear strains $\varepsilon_{13}$ and $\varepsilon_{23}$ and their influence on the core-energy and core-structure. In the present work, we do not study the influence of $\varepsilon_{12}$ strain, as, even at small values of shear stress resulting from this shear strain ($\sim 1.6~MPa$, cf.~\citet{Carter2013}), the edge dislocation glides by overcoming the Peierls barrier. The computed core-energy dependence on $\varepsilon_{13}$ and $\varepsilon_{23}$ shear strains is shown in figure~\ref{fig:shear_strain}. It is interesting to note that the core-energy dependence is symmetric in $\varepsilon_{13}$, to within the discretization errors of $0.005~eV$. On the other hand, the core-energy dependence on $\varepsilon_{23}$ strain is asymmetric. This qualitative change in core-energy dependence can be rationalized from the forces created by these macroscopic shear strains on the Shockley partials. The shear stress associated with $\varepsilon_{13}$ results in climb forces, in opposite directions, on the screw components of Shockley partials. Upon changing the sign of the shear stress, the direction of the climb forces on each of the Shockley partials is reversed, but these configurations are symmetry related and thus the core-energy variation is symmetric with respect to $\varepsilon_{13}$. In contrast, the shear stress associated with $\varepsilon_{23}$ results in glide forces, in opposite directions, on the screw component of Shockley partials. Thus, the sign of this shear stress has an effect of either increasing or decreasing the partial separation distance between the Shockley partials, which is the source of the asymmetry in core-energy dependence on $\varepsilon_{23}$. This is also further corroborated from the core-structure, which shows a significant change in the partial separation distance in the screw-component DD plot in response to $\varepsilon_{23}$ strain. In particular, the partial separation in the screw-component of the DD plot changes from $3.5|\mathbf{b}|$ at $\epsilon_{23} =-0.66$ to $5|\mathbf{b}|$ at $\epsilon_{23} =0.66$ (cf. figure~\ref{fig:DD_screw_e23}). Further, the partial separation in the edge-component of the DD plot also changes from $4.5|\mathbf{b}|$ at $\epsilon_{23} =-0.66$ to $5.5|\mathbf{b}|$ at $\epsilon_{23} =0.66$. On the other hand, the core-structures are found to be similar for both negative and positive $\varepsilon_{13}$ strains of a given magnitude, as expected from symmetry. For the range of $\varepsilon_{13}$ shear strains considered in the present work, the partial separation in the edge-component DD plots is in the range of $4.5-5~|\mathbf{b}|$ and the partial separation in the screw-component DD plots is in the range of $4-4.5~|\mathbf{b}|$.

\subsection{Core-force}

The study on the effect of macroscopic strains revealed a strong dislocation core-energy dependence on macroscopic deformations. Interestingly, the slope of the core-energy dependence on macroscopic strains is non-zero at zero strain. This suggests that the energetics of dislocation-cores play a non-trivial role in governing the behavior of dislocations even at small strains and at macroscopic scales. In order to elucidate the role of dislocation core-energy in influencing dislocation behavior, we consider the force on a unit length of dislocation line segment resulting from external loads or other defects. The classical force on the dislocation arising from elastic interactions is given by the Peach-Koehler force~\citep{Hirth,Peach-Koehler1950}:
\begin{equation}\label{eqn:PK}
\mathbf{f}_{PK} = \sigma\,.\,\mathbf{b}\times\xi\,,
\end{equation}
where $\sigma$ denotes the stress tensor, $\mathbf{b}$ denotes the Burgers vector and $\xi$ denotes the unit vector along the dislocation line. However, since the core-energy is dependent on macroscopic strains at the dislocation core, there is an additional configurational force associated with this dependence, referred to as the core-force, and is given by
\begin{equation}
\mathbf{f}_{core} = -\frac{\partial E_{core}}{\partial \varepsilon_{ij}} \nabla \varepsilon_{ij} \,.
\end{equation}
Given the non-linear nature of the core-energy dependence on macroscopic strains, $\frac{\partial E_{core}}{\partial \varepsilon_{ij}}$, in general, is a function of the macroscopic strain at the dislocation core. However, for typical strains of $<0.1\%$ occurring in most deformations in solids, $\frac{\partial E_{core}}{\partial \varepsilon_{ij}}$ can be computed, to leading order, from the slopes of the core-energy dependence on various uniaxial and shear strains at zero strain (cf. figures~\ref{fig:uniaxial1}-\ref{fig:uniaxial3},~\ref{fig:shear13}-\ref{fig:shear23}), i.e., $\frac{\partial E_{core}}{\partial \varepsilon_{ij}}\approx \frac{\partial E_{core}}{\partial \varepsilon_{ij}}\big{|}_{\varepsilon_{ij}=0}$. It is interesting to note that, while the Peach-Koehler force depends on the stress-tensor (which is related to the strain-tensor), the core-force depends on the gradient of strain-tensor and can be significant in regions of inhomogeneous deformations. We note that past studies (cf. e.g.~\citet{Gehlen1972,Henager2005,Clouet2011}), employing elastic formulations to model the defect-core using dislocation dipoles and line forces, have investigated the interaction of dislocation cores with applied pressure and have suggested corrections to the Peach-Koehler force based on these interactions. However, a direct quantification of this core-force solely from electronic-structure calculations has been beyond reach heretofore.

In order to understand the implications of the core-energy dependence on macroscopic strains on the behavior of dislocations, we consider the interaction between two dislocations. In particular, we consider a dislocation dipole, with one dislocation (dislocation $A$) along the $Z-$axis located at $(X,Y)=(0,0)$ with a Burgers vector ($\mathbf{b}$) of $\frac{a_0}{2}[110]$ and another dislocation (dislocation $B$), also along $Z-$axis, located at $(X,Y)=(x,y)$ with a Burgers vector of $-\frac{a_0}{2}[110]$ ($-\mathbf{b}$). The Peach-Koehler force~\citep{Peach-Koehler1950,Hirth} on a unit length of dislocation $B$ due to the elastic fields from dislocation $A$ is given by
\begin{equation}
\mathbf{f}^{B}_{PK}=\sigma^{A}.(-\mathbf{b})\times\xi^{B}\,,
\end{equation}
where $\sigma^A$ denotes the stress tensor associated with the elastic fields produced by dislocation $A$ at the location of dislocation $B$, and $\xi^{B}$ denotes the unit vector along the dislocation line of $B$ (unit vector along $Z-$axis). We note that the Peach-Koehler force ($\mathbf{f}_{PK}$) decays as $O(\frac{1}{r})$, where $r$ denotes the distance between the two dislocations, which follows from the asymptotic behavior of the elastic field of a dislocation. On the other hand, the core-force---force resulting from the macroscopic strain dependence of the core-energy---is given by
\begin{equation}
\mathbf{f}^{B}_{core}=-\nabla\big(E_{core}^{A}(\varepsilon^{B}_{ij})+E_{core}^{B}(\varepsilon^{A}_{ij})\big)\,,
\end{equation}
where $E_{core}^{A}(\varepsilon^{B}_{ij})$ denotes the core-energy of dislocation A in the elastic field of B ($\varepsilon^{B}_{ij}$), and $E_{core}^{B}(\varepsilon^{A}_{ij})$ denotes the core-energy of dislocation B in the elastic field of A ($\varepsilon^{A}_{ij}$). For the dipole configuration, we note that $\varepsilon^{B}_{ij}=\varepsilon^{A}_{ij}$~\citep{Hirth}, and, from symmetry, $E_{core}^{A}(\varepsilon^{B}_{ij})=E_{core}^{B}(\varepsilon^{A}_{ij})$. Thus,
\begin{equation}
\mathbf{f}_{core}^{B}=-2\,\nabla\,\big( E_{core}^{B}(\varepsilon^{A}_{ij})\big) = -2\frac{\partial E_{core}}{\partial {\varepsilon_{ij}}}\Big|_{\varepsilon_{ij}=0}\nabla{\varepsilon^{A}_{ij}}.
\end{equation}
Noting that $\nabla\varepsilon_{ij}$ is $O(\frac{1}{r^2})$, the core-force for dipole interactions decays as $O(\frac{1}{r^2})$. Thus, this is a shorter-ranged force in dislocation-dislocation interactions compared to the Peach-Koehler force.

\begin{figure}[h]
\centering
\includegraphics[width=1.0\textwidth]{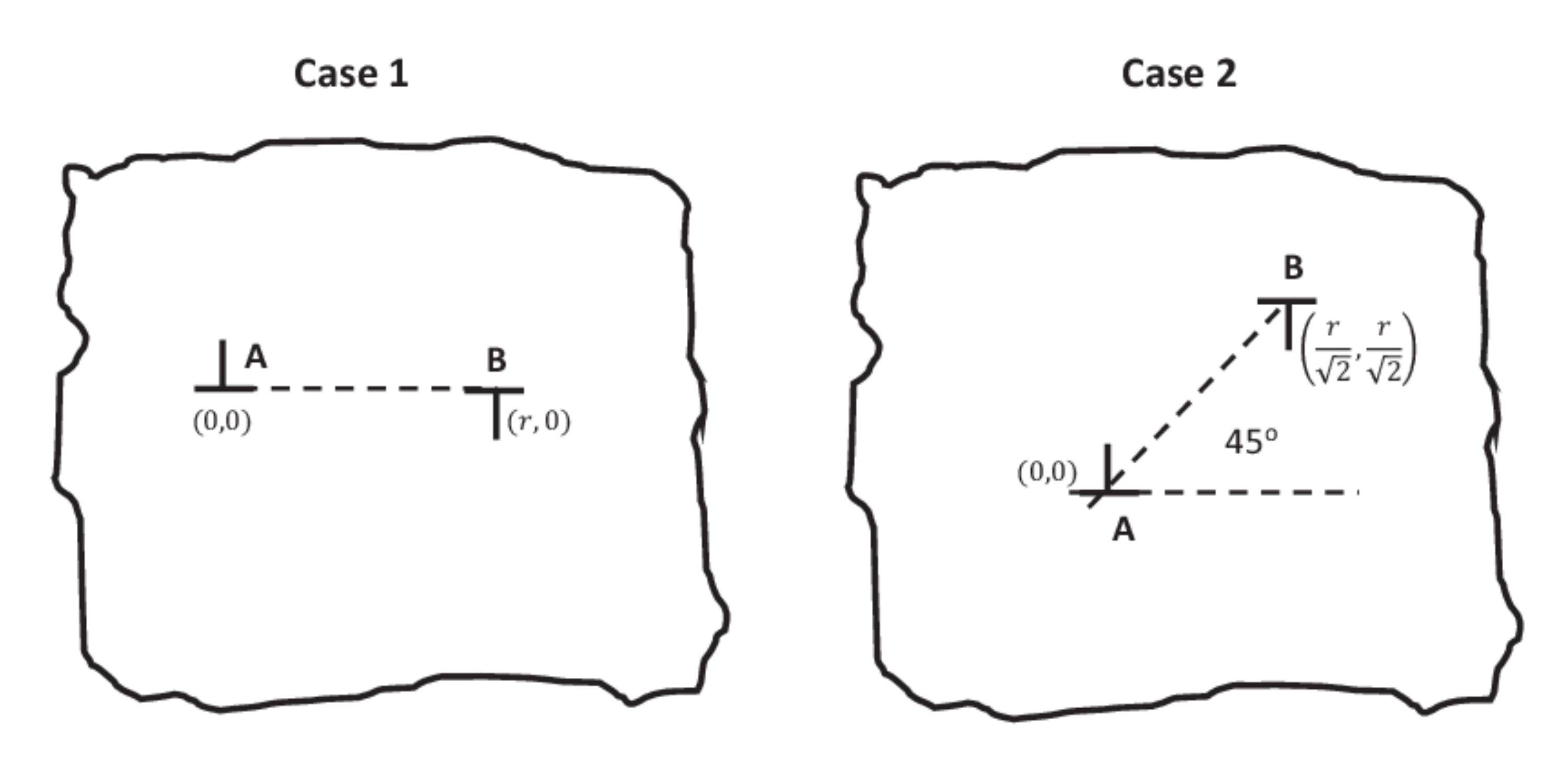}
\caption{Schematics for the two scenarios considered. \emph{Case (i)}: Edge dislocations aligned along the glide-plane. \emph{Case (ii)}: Dislocations at an angle of $45^\circ$ to the glide-plane.}
\label{fig:caseStudy}
\end{figure}

In order to further investigate the quantitative aspects of this core-force and its implications on dislocation-dislocations interactions, we consider two specific cases as shown in figure~\ref{fig:caseStudy}. In the first case, the dislocations are aligned along the glide-plane, whereas in the second case they are at an angle of $45^{\circ}$. For these two cases, we computed the Peach-Koehler force and the core-force on dislocation B.

\emph{Case (i):} For the first case, using the stress tensor associated with the elastic fields of a stationary edge dislocation~\citep{Hirth}, the Peach-Koehler force is given by
\begin{equation}
\mathbf{f}_{PK}^{B}=-\frac{Gb^2}{2\pi(1-\nu)}\frac{1}{r}\,\hat{\mathbf{i}} \,,
\end{equation}
where $\hat{\mathbf{i}}$, $\hat{\mathbf{j}}$, $\hat{\mathbf{k}}$ denote unit vectors along $X-Y-Z$ axis, and $b=|\mathbf{b}|$. The corresponding core-force, using the strain field of a stationary edge dislocation, is given by
\begin{equation}
\begin{split}
\mathbf{f}_{core}^{B}=&-2\frac{\partial E_{core}}{\partial {\varepsilon_{ij}}}\Big|_{\varepsilon_{ij}=0}\nabla{\varepsilon^{A}_{ij}}\\
=&\Big(\frac{\partial E_{core}}{\partial {\varepsilon_{11}}}\Big|_{\varepsilon_{ij}=0} \frac{2\lambda+3G}{\lambda+2G} - \frac{\partial E_{core}}{\partial {\varepsilon_{22}}}\Big|_{\varepsilon_{ij}=0} \frac{2\lambda+G}{\lambda+2G} \Big) \frac{b}{\pi r^2} \,\hat{\mathbf{j}} \,.
\end{split}
\end{equation}
The second equality in the above expression for core-force uses the fact $\nabla\varepsilon^{A}_{33}=\nabla\varepsilon^{A}_{23}=\nabla\varepsilon^{A}_{13}=0$ (from the strain field of a stationary edge dislocation), and $\frac{\partial E_{core}}{\partial \varepsilon_{12}}\big{|}_{\varepsilon_{ij}=0}=0$, from symmetry, for an edge dislocation. It is interesting to note that, for this case, while the Peach-Koehler force is a glide force on the dislocation, the core-force is a climb force.

\emph{Case (ii):} For the second case, the Peach-Koehler force and the core-force are given by
\begin{equation}
\mathbf{f}_{PK}^{B}=-\frac{Gb^2}{2\pi(1-\nu)}\frac{\sqrt{2}}{r}\,\hat{\mathbf{j}} \,,
\end{equation}
\begin{equation}\label{eqn:f_core_case2}
\begin{split}
\mathbf{f}_{core}^{B}=&-\Big(\frac{\partial E_{core}}{\partial {\varepsilon_{11}}}\Big|_{\varepsilon_{ij}=0} \, + \, \frac{\partial E_{core}}{\partial {\varepsilon_{22}}}\Big|_{\varepsilon_{ij}=0} \,\Big)\,\frac{G}{\lambda+2G} \,\frac{b}{\pi r^2}\, \hat{\mathbf{i}}\\
&-\Big(\frac{\partial E_{core}}{\partial {\varepsilon_{11}}}\Big|_{\varepsilon_{ij}=0}\, - \, \frac{\partial E_{core}}{\partial {\varepsilon_{22}}}\Big|_{\varepsilon_{ij}=0}\,\Big)\,\frac{\lambda+G}{\lambda+2G} \,\frac{b}{\pi r^2}\, \hat{\mathbf{j}}\,.
\end{split}
\end{equation}
The above expression for core-force uses the fact $\nabla\varepsilon^{A}_{33}=\nabla\varepsilon^{A}_{23}=\nabla\varepsilon^{A}_{13}=0$ (from the strain field of a stationary edge dislocation), and $\frac{\partial E_{core}}{\partial \varepsilon_{12}}\big{|}_{\varepsilon_{ij}=0}=0$ from symmetry. It is interesting to note that, for this case, while the Peach-Koehler force is a climb force on the dislocation, the core-force has both glide and climb components.

We next investigate the relative importance of the core-force in governing the dislocation behavior. To this end, we estimated the glide component of the core-force for \emph{Case (ii)} from equation~\eqref{eqn:f_core_case2}, and compared it to the Peierls-Nabarro force---the critical force that will cause dislocation glide. For these calculations, we used the computed elastic constants $\{G,\nu,\lambda\}=\{29\,GPa, 0.31, 47.3\,GPa\}$. Using $1.6~MPa$ for the Peierls stress~\citep{Carter2013}, we computed the Peierls-Nabarro force (from equation~\ref{eqn:PK}) on a $1~\mbox{\AA}$ length of dislocation line to be $2.85\times10^{-5}$ $eV/\mbox{\AA}$. Using $\frac{\partial E_{core}}{\partial {\varepsilon_{11}}}\Big|_{\varepsilon_{ij}=0}=-9.10\,eV/\mbox{\AA}$ and $\frac{\partial E_{core}}{\partial {\varepsilon_{22}}}\Big|_{\varepsilon_{ij}=0}=16.34\,eV/\mbox{\AA}$, estimated from the data in figures~\ref{fig:uniaxial1} and \ref{fig:uniaxial2}, the glide component of the core-force from case (ii) on a $1~\mbox{\AA}$ length of dislocation line is computed to be $\frac{-1.82}{r^2}$ $eV/\mbox{\AA}$ where $r$ is the distance between the dislocations $A$ and $B$ measured in Angstroms. It is interesting to note that even at a distance of $25~nm$, the magnitude of core-force is larger than the magnitude of Peierls-Nabarro force. These results suggest that the short-ranged core-force, resulting from the dependence of core-energy on macroscopic deformations, is significant up to tens of nanometers and can play an important role in governing the behavior of dislocations, especially in regions of inhomogeneous deformations.

\section{Conclusions}\label{sec:Conclusions}
In the present work, we employed a real-space formulation of orbital-free DFT with finite-element discretization to study an edge dislocation in Aluminum. The local real-space formulation and the finite-element discretization are crucial in accurately quantifying the energetics of an isolated dislocation, which is otherwise not accessible using a Fourier-space based formulation. In this study, we computed the core-size of an edge dislocation by identifying the region where the perturbations in the electronic-structure arising from the defect-core, which can not be accounted for in continuum theories, are significant and have a non-trivial contribution to the dislocation energy. This
allows us to unambiguously identify the core-size of an isolated dislocation from the viewpoint of energetics. We estimated the core-size of an edge dislocation in Aluminum to be about $10|\mathbf{b}|$, which is much larger than conventional core-size estimates of $1-3~|\mathbf{b}|$. The relaxed core-structure showed two Shockley partials with a partial separation distance of $12.8~\mbox{\AA}$ that is consistent with prior electronic-structure studies of an edge dislocation in Aluminum. The core-energy---energy of relaxed Shockely partials corresponding to a core-size of $10|\mathbf{b}|$---per unit length of dislocation is computed to be $0.4~eV/\mbox{\AA}$.

We subsequently investigated the effect of macroscopically applied deformations on the core-energy and core-structure of an isolated edge dislocation. Interestingly, our studies suggest that the dislocation core-energy is significantly influenced by the applied macroscopic deformations. This core-energy dependence on macroscopic deformations results in an additional configurational force on the dislocation, beyond the Peach-Koehler force, and is referred to as \emph{core-force} in this work. This core-force is proportional to the gradient of the strain-tensor and can be significant in regions of inhomogeneous deformations. We quantified this core-force for a dislocation dipole for two specific configurations of the dipole: (i) dislocations aligned along the slip plane; (ii) dislocations at a $45^\circ$ angle to the slip plane. Interestingly, for the first configuration, where the Peach-Kohler force is a glide force on the dislocation, the core-force is a climb force on the dislocation. For the second configuration, while the Peach-Koehler force is a climb force on the dislocation, the core-force has both glide and climb components on the dislocation. We estimated the glide component of the core-force for the second configuration and compared it to the Peierls-Nabarro force, which is the critical force required for dislocation glide. Surprisingly, even up to $25~nm$ the glide component of the core-force was greater than the Peierls-Nabarro force, suggesting that the core-force can play an important role in governing dislocation behavior, possibly even at mesoscopic scales. Investigations on the influence of macroscopic deformations on the core-structure revealed that the separation between the Shockley partials is only marginally influenced by uniaxial, biaxial and triaxial strains. However, the core-structure is significantly influenced by the shear strain in the plane containing the dislocation line and the direction normal to the slip plane. This is expected as the resulting shear stress exerts a glide force on the screw-component of the partials, subsequently leading to a significant change in the partial separation.

The present work has provided the framework for accurately quantifying the energetics of isolated dislocations using electronic-structure calculations. While the present study focussed on an edge dislocation in Aluminum, concurrent efforts are also focussed on studying a screw dislocation. Further, ongoing and future efforts are also aimed at extending the computational methods to enable similar studies using Kohn-Sham DFT, which is an important step towards considering materials systems where available orbital-free kinetic energy functionals do not offer the desired accuracy. Investigating the role of core-force in governing the mechanical response at the mesoscopic scale, possibly using dislocation dynamics simulations, also presents an important direction for future research.

\section*{Acknowledgments}
We thank Dr. Roman Gr$\ddot{o}$ger for assistance with the differential displacement plots of Shockley partials. We are grateful to the support of National Science Foundation (Grant number CMMI0927478) under the auspices of which part of the work was conducted. The study on the effect of macroscopic deformations on dislocation energetics was conducted as part of research supported by the US Department of Energy, Office of Basic Energy Sciences, Division of Materials Science and Engineering under award number DE-SC0008637 that funds the PRedictive Integrated Structural Materials Science (PRISMS) Center at University of Michigan. V. G. gratefully acknowledges the support of Alexander von Humboldt Foundation through a research fellowship, and is grateful to the hospitality of the Max-Planck Institute for Mathematics in Sciences while completing this work.


\end{document}